\documentclass{article}
\title{Using joint models in phase I dose-finding designs in oncology: considerations for frequentist approaches}
\usepackage{graphicx}
\usepackage{color}
\usepackage{floatrow}
\newfloatcommand{capbtabbox}{table}[][\FBwidth]
\usepackage{algpseudocode}
\usepackage{authblk}
\author[1]{Xijin Chen}
\author[1]{Pavel Mozgunov}
\author[2]{Richard D. Baird}
\author[1,3]{Thomas Jaki}
\affil[1]{MRC Biostatistics Unit, University of Cambridge, Cambridge, UK}
\affil[2]{Cancer Research UK, Cambridge Centre, University of Cambridge, Robinson Way, Cambridge, UK}
\affil[3]{University of Regensburg, Germany}
\usepackage{marvosym}
\usepackage{caption}
\usepackage{colortbl}
\usepackage{color,soul}
\usepackage{multirow}
\usepackage[toc,page]{appendix}
\usepackage[backend=biber, sorting=none,style=numeric,citestyle=numeric]{biblatex}
\usepackage{longtable}
\usepackage{booktabs}
\usepackage{makecell}
\usepackage{outlines}
\usepackage{verbatim}
\usepackage[parfill]{parskip}
\usepackage{setspace}
\usepackage{float}
\usepackage{graphicx}
\usepackage{xcolor}
\usepackage{soul}
\usepackage{enumitem}
\usepackage{hyperref}
\usepackage{import}
\usepackage{listings}
\usepackage{authblk}
\usepackage[margin=1in]{geometry}
\usepackage{enumitem}

\usepackage{subcaption}
\usepackage{amsmath}

\setul{0.5ex}{0.3ex}
\definecolor{Red}{rgb}{1,0.0,0.0}
\setulcolor{Red}

\begin{document}
\maketitle
\begin{abstract}
Dose-finding trials for oncology studies are traditionally designed to assess safety in the early stages of drug development. With the rise of molecularly targeted therapies and immuno-oncology compounds, biomarker-driven approaches have gained significant importance. In this paper, we propose a novel approach that incorporates multiple values of a predictive biomarker to assist in evaluating binary toxicity outcomes using the factorization of a joint model in phase I dose-finding oncology trials. The proposed joint model framework, which utilizes additional repeated biomarker values as an early predictive marker for potential toxicity, is compared to the likelihood-based continual reassessment method (CRM) using only binary toxicity data, across various dose-toxicity relationship scenarios. Our findings highlight a critical limitation of likelihood-based approaches in early-phase dose-finding studies with small sample sizes: estimation challenges that have been previously overlooked in the phase I dose-escalation setting. We explore potential remedies to address these challenges and emphasize the appropriate use of likelihood-based methods. Simulation results demonstrate that the proposed joint model framework, by integrating biomarker information, can alleviate estimation problems in the the likelihood-based continual reassessment method (CRM) and improve the proportion of correct selection. However, we highlight that the inherent data limitations in early-phase dose-finding studies remain a significant challenge that cannot fully be overcomed in the frequentist framework.
\\


\end{abstract}
\section{Introduction}

Dose-finding studies, particularly phase I trials, primarily aim to assess toxicity before evaluating efficacy and comparing new treatments with a standard-of-care in subsequent phases. These trials are traditionally designed for cytotoxic agents, being based on tumor size and histology. The advent of molecularly targeted therapies and immuno-oncology compounds has revolutionized oncology drug development~\parencite{spreafico2021future}. Regulatory agencies like the Food and Drug Administration (FDA) champion innovations of modern approaches, such as master protocol, adaptive designs, and biomarker-driven evaluations for patient selection and endpoint determination~\parencite{FDAinno}. Initiatives like Project Optimus highlight the ongoing transformation of dose optimization and selection paradigms~\parencite{FDAOptimus}, enabling better dose characterization.

Biomarkers, which are defined as objectively measurable indicators of biological or pharmacological processes, play a central role in the evolution of paradigm for dose selection~\parencite{biomarkers2001biomarkers}. %Among the seven categories of biomarkers defined by the BEST glossary, predictive biomarkers are particularly valuable for forecasting treatment response~\parencite{FDAbiomarker}. 
Modern technologies such as liquid biopsies provide a minimally invasive approach to capturing the comprehensive genetic landscape of tumors in real time throughout the course of treatment~\parencite{kilgour2020liquid}. Among these, circulating tumor DNA (ctDNA) has shown considerable promise for monitoring therapeutic response, identifying resistance mechanisms, and predicting early relapse~\parencite{yates2015subclonal, shah2009mutational}. Similarly, platelet counts have emerged as valuable biomarkers for assessing disease recurrence and forecasting treatment-related toxicity or adverse events~\parencite{gresele2024low, huplatelet}. %In parallel, ongoing efforts aim to establish standardized ctDNA response criteria to support its integration into clinical practice~\parencite{jakobsen2023ctdna}. 
These developments underscore the potential of using predictive biomarkers, particularly in dose-finding studies, where it can provide real-time assessments of treatment-related outcomes. By enabling precise and dynamic monitoring of tumor behavior, biomarkers such as platelet counts could revolutionize treatment strategies and improve dose strategy in phase I trials in oncology. 
%Despite these advances, the integration of serial biomarker measurements into early-phase clinical trials remains underexplored, particularly in the context of optimizing dose selection. Addressing this gap, this study investigates the utility of serial measurements of generic biomarkers to improve predictive accuracy and improve real-time monitoring for optimized dose strategy in phase I clinical trials.  
Notably, persistently low or declining platelet counts during treatment may serve as early indicators of increasing risk, while stable or increasing levels may reflect a more favorable safety profile~\parencite{gresele2024low}. Advances in next-generation sequencing (NGS) technologies can substantially enhance the qualification and reliability of the use of these predictive biomarkers~\parencite{satam2023next}.
%~\parencite{warburton2023detectable,vo2025circulating}
%For example, there is an overview of available data on circulating biomarkers of response and toxicity of immunotherapy in lung cancer 

%patients~\parencite{indini2021circulating}. 
%Analytical and clinical validation are always required before a biomarker can be used in clinical settings. Certain biological characteristics of ctDNA remain unknown~\parencite{ou2021biomarker,sanchez2022circulating}. It is necessary to capture a treatment's effect on clinically relevant outcomes as substitutes for a surrogate endpoint, whereas the level of evidence required for a predictive biomarker is less stringent. Predictive biomarkers are particularly suitable for providing actionable information during early phase development without requiring the same level of strict validation as surrogate biomarkers~\parencite{wickstrom2017biomarkers}.

Incorporating a generic biomarker as a predictive marker into dose-finding designs necessitates the joint modeling of mixed outcomes measured on different scales. For example, Chen et al.~\parencite{chen2025using} incorporate ctDNA as a predictive biomarker for efficacy by dichotomizing the continuous biomarker based on a threshold value, thereby facilitating its integration into the trial design. However, this dichotomization approach may result in a loss of information, potentially reducing the statistical efficiency and predictive power of the biomarker.

Traditional dose-finding studies have primarily aimed to identify the maximum tolerated dose (MTD) based on dose-limiting toxicities (DLTs)—serious adverse events that impair normal activities and require therapeutic intervention~\parencite{jaki2013principles}. However, with many novel oncology agents, DLTs may be infrequent or absent, prompting a shift toward biomarker-driven dose escalation strategies. While DLTs are often analyzed as a binary endpoint in dose-finding studies, biomarkers such as platelets are repeatedly measured on a continuous scale. Standard multivariate tools suffice for outcomes measured on the same scale, but jointly analyzing non-commensurate outcomes requires specialized statistical methods~\parencite{teixeira2013factorization}. There are basically two classes of general likelihood-based approaches to jointly model binary and continuous outcomes~\parencite{teixeira2009correlated}. The first class considers a factorization of the joint likelihood and fits a model to each component of the factorization~\parencite{catalano1992bivariate,cox1992response,fitzmaurice1995regression}. The other class uses a latent variable shared by both outcomes inducing the correlation~\parencite{sammel1997latent,dunson2000bayesian}. 
%~\parencite{de2013analysis}

The concept of factorizing a joint model into a marginal model and a conditional model for different components of mixed outcomes is particularly useful in dose-finding studies, especially when the mixed outcomes are not always observed simultaneously. Bekele and Shen~\parencite{bekele2005bayesian} proposed a joint model for binary toxicity and continuous activity outcomes by factorizing the joint distribution into a marginal model for the continuous outcome and a conditional model for the binary outcome, assuming both outcomes are evaluated at the same time. Wason and Seaman~\parencite{wason2013using} proposed a method for making inferences on composite or mixed efficacy outcomes by fitting a joint model to the continuous and binary efficacy components in phase II cancer studies. This was further expanded to a multivariate probit model incorporating toxicity outcomes in seamless phase I/II studies~\parencite{wason2020latent}. However, neither approach has been comprehensively evaluated in adaptive dose-finding studies involving multiple continuous outcomes assessed over varying evaluation windows. Our proposed method addresses this gap by accommodating continuous biomarkers—such as platelet counts—measured on different time scales, thereby enhancing the flexibility and applicability of joint modeling in complex clinical trial settings. Specifically, we focus on phase I dose-finding trials, modeling both toxicity and predictive biomarkers associated with toxicity outcomes.

%Zeger and Liang~\parencite{zeger1986longitudinal} have discussed factorization approaches to modelling mixed outcomes in the context of longitudinal studies.
\par

Among model-based approaches for dose-finding designs, the continual reassessment method (CRM) stands as one of the most widely recognized model-based phase I trial designs~\parencite{o1990continual}. Originally developed within a Bayesian framework, it was later adapted into a frequentist variant known as the CRM likelihood approach (CRML)~\parencite{o1996continualcrml}. The CRML method has the notable advantage of eliminating the need for prior specification. Further investigations via simulation studies into both Bayesian and likelihood-based CRM methods followed~\parencite{paoletti2009comparison}. 

%The CRM and CRML were originally developed to identify the maximum tolerated dose (MTD) with a primary focus on safety, which can be easily to dose-finding studies with efficacy outcomes, thus shifting the focus toward efficacy-driven dose escalation. In such cases, the primary objective becomes identifying the minimum efficacious dose (MED).

%for Phase I trials to identify the maximum tolerated dose (MTD) with a primary focus on safety, CRM can also be effectively adapted for dose-finding studies aimed at identifying the minimum effective dose (MED) by evaluating efficacy. In this work, we employ a CRM-based approach using the probit model as the foundational baseline. 

%Building on the factorization framework, this study investigates the incorporation of ctDNA as a biomarker within a joint model to enhance dose-finding designs.

% However, for many novel oncology agents, DLTs may not occur at all 
 
Our work introduces a factorization framework that incorporates repeated biomarker measurements taken at different time points along with binary toxicity outcomes in phase I dose-finding studies. We compare two variants of the proposed joint method: the Joint2d method, which utilizes a single biomarker measurement, and the Joint9d method, which integrates multiple biomarkers collected across multiple time points. These methods are compared with the traditional probit method, which is based solely on binary toxicity outcomes. Our findings aim to demonstrate the potential of biomarker to refine dose-finding methodologies in oncology, particularly by improving the accuracy of dose selection by including a repeatedly measured biomarker (such as platelet counts). 

In the following sections, we outline the factorization framework in Section~\ref{sec:methods}, illustrate challenges in parameter estimation in Section~\ref{sec:problem}, propose different remedies in Section~\ref{sec:remedies}, and present simulation results in different scenarios in Section~\ref{sec:simulation} to highlight the advantages of incorporating biomarker in phase I dose-finding studies in oncology. The paper concludes with a discussion presented in Section~\ref{sec:discussion}.

\section{Methods}\label{sec:methods}

%This section starts by introducing the notation and settings in Section~\ref{subsec:notation}, followed by a description of the dose-finding algorithm in Section~\ref{subsec:algorithm} and the introductino of notations for ctDNA~. Next, we present the general framework based on factorization, starting with the Probit method based on binary efficacy outcomes only. Building on this foundation, we extend the framework to joint methods that incorporate binary outcomes with varying numbers of continuous ctDNA measurements.

\subsection{Model-based dose-finding designs based on the binary outcome only}\label{subsec:notation}

We consider a dose-finding trial with $J$ doses under investigation, notated as $d_1,...,d_J$. Each patient enrolled in the trial is assessed for a binary toxicity outcome, denoted as $Y_b$, which indicates the presence or absence of dose-limiting toxicity (DLT) at those doses.

%and (2) a vector of continuous ctDNA measurements, denoted as $\mathbf{Y_{c}}$, recorded weekly over a full treatment cycle. 

%By incorporating this correlation, the model integrates both weekly molecular dynamics (via ctDNA) and treatment outcomes (via binary efficacy outcome).

To model the dose-toxicity relationship, a working model $\psi(x, \beta)$ can be employed, where $\beta$ parameterizes the relationship. An important point about the CRM is that the \text{dose labels} are not the actual doses administered, but rather are defined on a conceptual scale that represents an ordering of the risks of events, termed as the skeleton~\parencite{cheung2011dose}. The impact of dose mapping on the performance of the CRM has been evaluated~\parencite{garrett2006continual,cheung2002simple}. `Backward fitting', which is widely accepted in the CRM literature, maps the actual doses from the the numerical dose levels through the functional form that best mirrors that shape. That is, dose labels $x$ are transformed from skeleton using backward substitution into the model. Specifically, these labels are derived by solving $x = {\psi}^{-1}(p_{0}, \beta^*)$, where $\beta^*$ represents the initial guess of the parameter $\beta$, and $p_{0}$ corresponds to the skeleton probabilities for the investigated doses. This transformation ensures that the model aligns accurately with the guess about toxicity probability estimates provided by clinicians, creating a strong foundation for dose-toxicity evaluation. 

%By fitting the skeleton, the working model $\psi(x, \beta)$ achieves a consistent representation of expected response probabilities at each dose level.

Table~\ref{tab:ModelDose} presents examples of candidate models $\psi(x, \beta)$ and the corresponding dose labels $x_j$ associated with the actual doses $d_j$ in the set ${d_1, \dots, d_J}$. Further details regarding model choice and the design parameters of the CRM approach can be found in Wheeler et al.~\parencite{wheeler2019design}. To ensure an increasing dose-toxicity relationship, these models use $\exp(\beta_{b1})$, as shown in Table~\ref{tab:ModelDose}.

\begin{table}[!ht]
\centering
\begin{tabular}{lllll}
\hline
\multirow{2}{4em}{Model} & $\psi(x, \beta)$ & {Dose labels $x_j$} & {Choice of $\beta^*$ } & {Dose labels $x_j$} \\
 & (General form ) & (initial guess) & (given $\beta^*$) \\
\hline
Power (Empiric)  & $x^{\text{exp}(\beta_{b1})}$ & $p_j^{\frac{1}{\exp(\beta_{b1})}}$ & $\beta_{b1} = 0$ & $p_j$ \\
%One-parameter logistic & $\frac{\text{exp}(a_0+\text{exp}(\beta_{b1})x)}{1+\text{exp}(a_0 +\text{exp}(\beta_{b1})x)}$  & $\frac{\ln\left(\frac{p_j}{1 - p_j}\right) - a_0}{\text{exp}(\beta_{b1})}$ & $\beta_{b1} = 0$ & $\ln\left(\frac{p_j}{1 - p_j}\right)-a_0$ \\
%Two-parameter logistic & $\frac{\text{exp}(\beta_{b0}+\text{exp}(\beta_{b1})x)}{1+\text{exp}(\beta_{b0}+\text{exp}(\beta_{b1})x)}$ & $\frac{\ln\left(\frac{p_j}{1 - p_j}\right) - \beta_{b0}}{\text{exp}(\beta_{b1})}$ & $\beta_{b0} = 0,\ \beta_{b1} = 0$ & $\ln\left(\frac{p_j}{1 - p_j}\right)$ \\
\hline
One-parameter probit & $\Phi(a_0+\text{exp}(\beta_{b1})x)$   & $\frac{\Phi^{-1}(p_j)-a_0 }{\text{exp}(\beta_{b1})}$ & $\beta_{b1} = 0$ & $\Phi^{-1}(p_j)-a_0 $ \\
\hline
Two-parameter probit  & $\Phi(\beta_{b0}+\text{exp}(\beta_{b1})x)$ & $\frac{\Phi^{-1}(p_j)-\beta_{b0}}{\text{exp}(\beta_{b1})}$  & $\beta_{b0} = 0,\ \beta_{b1} = 0$ & $\Phi^{-1}(p_j)$ \\
\hline
\end{tabular}
\caption{Common choices for dose-toxicity models and resultant dose labels for the CRM. $p_j$ is the skeleton probability at the dose label $x_j$ and $a_0$ is the fixed intercept in corresponding one-parameter models.} 
\label{tab:ModelDose}
\end{table}

\subsection{Dose-finding algorithm: The two-stage design}\label{subsec:algorithm}

The dose-finding algorithm follows that of the CRML~\parencite{o1996continualcrml} that focuses on the evaluation of toxicity. The CRML utilizes that benefit of the two-stage design first proposed by Storer~\parencite{storer1989design} in the context of the classical up-and-down schemes and limits their attention to the power model. Specifically, the design comprises two distinct phases: an initial exploratory escalation stage (Stage 1) and a subsequent modeling stage (Stage 2) for refining dose selection. These two phases operate independently of one another.

In Stage 1, the heterogeneity requirement is imposed, necessitating the observation at least one DLT and one non-DLT~\parencite{o1996continualcrml}. This stage plays a critical role in the context of sequential likelihood estimation, as the likelihood equation remains monotonic in the unknown parameter until heterogeneity in responses is observed. Meeting this requirement was intended to ensure that the maximization of the log-likelihood occurs within the interior of the parameter space, thereby enabling more accurate parameter estimation. 

%In the original CRML framework, this requirement was addressed by assuming that the first three outcomes (i.e., the first cohort) at the lowest dose level were non-toxicities, thereby ensuring the continuity of the trial.

The requirement for heterogeneity is well-established in the literature and originates from the CRML approach to ensure the existence of a maximum likelihood estimate (MLE) for the chosen power model. This requirement originates from more general situations and richer models of two-parameter logistic model, which has been extensively studied in Silvapulle~\parencite{silvapulle1981existence}. Albert and Anderson~\parencite{albert1984existence} also identified a certain degree of overlap in the data as a necessary and sufficient condition for the MLE to exist. Similarly, Lesaffre and Kaufmann~\parencite{lesaffre1992existence} established a corresponding condition for the probit model, which aligns with the logistic model framework. This condition, known as the \textit{separation condition}, requires a degree of intermixed or overlapping data points to ensure MLE existence in both logistic and probit models. 

In binomial response models, data configurations are mutually exclusively categorized as complete separation, quasi-complete separation, or overlap. Table~\ref{tab:dataconfig} presents a 2 $\times$ 2 tabular representation of the data configurations, specifically tailored to the initial stage starting from the lowest dose in phase I dose-finding context. This representation focuses on the two lowest dose levels to reflect conditions typically encountered at the beginning of such studies.

%Figure 1 in Schwendinger et al.~\parencite{schwendinger2021comparison} offers a clear illustration of these three cases, highlighting the patterns of data points in the observation space. 
\begin{table}[h]
    \begin{subtable}[h]{0.32\textwidth}
        \centering
        \begin{tabular}{ccc}
        \hline
         & DLTs & Non-DLTs \\
        \hline 
        $d_1$ & 0 & 3\\
        $d_2$ & 3 & 0\\
        \hline
       \end{tabular}
       \caption{Complete separation}
    \end{subtable}
    \hfill
    \begin{subtable}[h]{0.32\textwidth}
        \centering
        \begin{tabular}{ccc}
        \hline
         & DLTs & Non-DLTs \\
        \hline 
        $d_1$ & 0 & 3 \\
        $d_2$ & 3 & 3 \\
        \hline
        \end{tabular}
        \caption{Quasi-complete separation}
     \end{subtable}
     \hfill
     \begin{subtable}[h]{0.32\textwidth}
        \centering
        \begin{tabular}{ccc}
        \hline
         & DLTs & Non-DLTs \\
        \hline 
        $d_1$ & 3 & 3\\
        $d_2$ & 3 & 3\\
        \hline
        \end{tabular}
        \caption{Overlap}
     \end{subtable}
     \caption{Three mutually exclusive binary outcome configurations at the initial stage of dose-finding studies with two dose levels.}
     \label{tab:dataconfig}
\end{table}

%In the original CRML design, the initial stage required data from at least two dose levels, with the first cohort assumed to demonstrate three non-toxic responses to ensure continuity in the trial. This requirement also prevents the fitting of a dose-response model based solely on data from a single dose level. Similarly, we require that the initial stage include at least two cohorts, each assigned to different dose levels. This aligns with the original proposal of CRML and the principle of dose-finding studies, which attempt to evaluate multiple dose levels to accurately characterize the dose-response relationship. 

In light of the heterogeneity requirement incorporated into the CRML approach, the dose-finding algorithm is structured accordingly. Stage 1 consists of an initial phase before the implementation of the model in Stage 2. Corresponding flowchart is presented in Appendix~\ref{sup:flowchart}. The target dose is identified as the dose with the estimated probability closest to the target toxicity probability $\phi_T$, given by $d^T=\text{argmin}_{j \in \{1,...,J\}}|\psi(x^T,\hat\beta)-\phi_T|$. While the target toxicity level is typically specified by clinicians, an alternative approach involves referencing response rates from related treatments, falls outside the scope of the present work~\parencite{wheeler2019design}.

\begin{enumerate}
    \item \textbf{Initial Stage (Stage 1):}
    \begin{itemize}
        \item Administer the lowest dose level, $d_1$, to the first cohort of patients
        \item Evaluate whether the heterogeneity requirement is met based on the observed outcomes. If the criterion is satisfied, proceed to Stage 2
        \item If the criterion is not met, determine the next dose based on observed outcomes in the current cohort:
        \begin{itemize}
\item 0 DLTs
\begin{itemize}
\item If the current dose level is not the highest (\( d < 5 \)), escalate to the next higher dose.
\item If at the highest dose (\( d = 5 \)), remain at the same dose level
\end{itemize}
\item 1 DLT:
\begin{itemize}
\item Stay at the same dose level
\end{itemize}
\item \( \geq 2 \) DLTs:
\begin{itemize}
\item If at the lowest dose level (\( d = 1 \)):
\begin{itemize}
\item Apply the stopping rule ($\text{Pr}(p_T>0.3)\ge0.95$) if at least two DLTs occur at $d_1$, using all accumulated data to assess early stopping
\item If not, remain at the same dose level
\end{itemize}
\item If not at the lowest dose, de-escalate to the next lower dose level
\end{itemize}
\end{itemize}
    \end{itemize}
    \item \textbf{Modeling Stage (Stage 2):}
     \begin{itemize}
        \item Utilize Maximum Likelihood Estimation (MLE) with the selected model (e.g., the two-parameter probit model) to iteratively adjust dose administration, optimizing parameter estimates for the probability of toxicity
        \item Identify the target dose and determine the subsequent dose for administration, adhering to the no-skipping rule
        \item Conclude the trial upon reaching the pre-defined sample size, with the recommended dose $d^T$ determined based on the accumulated data
    \end{itemize}
\end{enumerate}

%Similarly, in our dose-finding design, the initial stage must establish heterogeneity between both dose levels and binary outcomes to facilitate reliable parameter estimation. 

%Importantly, this initial stage is independent of the ctDNA values, ensuring its consistency over a diverse range of outcomes. 

For both the joint method and the probit method, there is no requirement on biomarker outcomes in Stage 1. This implies that the consistent structure in the initial stage facilitates comparability across different modeling approaches.

\subsection{Notation}\label{subsec:notationctDNA}

While changes from baseline are commonly used to evaluate biomarkers in clinical practice, absolute measures—such as fragment length, variant allele frequency (VAF), and specific genomic alterations—are also frequently assessed. It has been shown that certain biomarker levels depend strongly on the cancer type. For simplicity, we denote the repeatedly measured biomarker as $\mathbf{Y}c = (Y{c1}, Y_{c2}, \ldots, Y_{ct})^T$, representing measurements taken at time points 1 through $t$ within a treatment cycle (e.g., week 1 to week $t$).

We assume a sequence of linearly decreasing mean biomarker values across dose levels, denoted as $\mu_{c1} > \mu_{c2} > \ldots > \mu_{ct}$, while maintaining a constant variance $\sigma_c$. This trend reflects the underlying assumption that lower biomarker levels are associated with an increased risk of adverse effects or toxicity, as exemplified by the relationship between declining platelet counts and treatment-related toxicity~\parencite{gresele2024low}. To capture the temporal relationship between the variances of consecutive biomarker measurements, we assume that the measurements taken closer in time are more strongly correlated in their predictive value. This temporal dependency is modeled using a first-order autoregressive process, AR(1), characterized by a variance-covariance matrix $\Sigma_c$. This assumption allows for simplicity in number of parameters. The correlation between biomarker measurements decays exponentially as the time gap between them increases, reflecting the sustained effects of treatment over time. For two time points, $t$ and $t+k$, the correlation is given by $\rho_c^k$, where $\rho_c$ ($0 < \rho_c < 1$) represents the autoregressive coefficient. Equation~\ref{eq:sigmac} illustrates $\Sigma_c$ using $t = 8$ as an example.

\begin{equation}
\Sigma_c=
\begin{pmatrix}
 \sigma_c^2 & \rho_c\sigma_c^2 & \rho_c^2\sigma_c^2 &  \rho_c^3\sigma_c^2  &  \rho_c^4\sigma_c^2 &  \rho_c^5\sigma_c^2  &  \rho_c^6\sigma_c^2  &  \rho_c^7\sigma_c^2 \\
  \rho_c\sigma_c^2  & \sigma_c^2 & \rho_c\sigma_c^2 & \rho_c^2\sigma_c^2 &  \rho_c^3\sigma_c^2  &  \rho_c^4\sigma_c^2 &  \rho_c^5\sigma_c^2  &  \rho_c^6\sigma_c^2 \\
  \rho_c^2\sigma_c^2 & \rho_c\sigma_c^2 & \sigma_c^2 & \rho_c\sigma_c^2  & \rho_c^2\sigma_c^2 &  \rho_c^3\sigma_c^2  &  \rho_c^4\sigma_c^2 &  \rho_c^5\sigma_c^2  \\ 
   \rho_c^3\sigma_c^2 &  \rho_c^2\sigma_c^2 & \rho_c\sigma_c^2  & \sigma_c^2 & \rho_c\sigma_c^2  & \rho_c^2\sigma_c^2 &  \rho_c^3\sigma_c^2  &  \rho_c^4\sigma_c^2   \\
   \rho_c^4\sigma_c^2 &  \rho_c^3\sigma_c^2 &  \rho_c^2\sigma_c^2 & \rho_c\sigma_c^2  & \sigma_c^2 & \rho_c\sigma_c^2  & \rho_c^2\sigma_c^2 &  \rho_c^3\sigma_c^2  \\
   \rho_c^5\sigma_c^2 &   \rho_c^4\sigma_c^2 &  \rho_c^3\sigma_c^2 &  \rho_c^2\sigma_c^2 & \rho_c\sigma_c^2  & \sigma_c^2 & \rho_c\sigma_c^2  & \rho_c^2\sigma_c^2 \\
   \rho_c^6\sigma_c^2 & \rho_c^5\sigma_c^2 & \rho_c^4\sigma_c^2 &  \rho_c^3\sigma_c^2 &  \rho_c^2\sigma_c^2 & \rho_c\sigma_c^2  & \sigma_c^2 & \rho_c\sigma_c^2 \\
  \rho_c^7\sigma_c^2  & \rho_c^6\sigma_c^2   & \rho_c^5\sigma_c^2 &   \rho_c^4\sigma_c^2 &  \rho_c^3\sigma_c^2 &  \rho_c^2\sigma_c^2 & \rho_c\sigma_c^2  & \sigma_c^2 \\
\end{pmatrix}.
\label{eq:sigmac}
\end{equation}

\subsection{Framework of factorization}\label{subsec:Framework}

For clarity, we demonstrate the framework using a bivariate example involving one binary variable $Y_b$ and one continuous variable $Y_c$. The key idea is to factorize the joint distribution and fit a univariate model to each component. We consider the factorization of a marginal model for continuous outcomes and a conditional model for the binary outcome~\parencite{de2013analysis}. This choice aligns with our context, where biomarker values can be assessed earlier than the binary toxicity outcomes. That is, the joint model of $Y_b$ and $Y_c$ is factorized in the form $f_{Yb,Yc}(y_b, y_c) = f_{Yc}(y_c)f_{Y_b|Y_c}(y_b|y_c)$. This factorization
can be motivated by assuming that the binary outcome $Y_b$ is a dichotomization of an underlying latent continuous variable $Y_b^*$.

\begin{equation}
Y_{b}=\left\{\begin{array}{ll}
0,&\text{ if } Y_{b}^* \leq 0 \\
1,&\text{ if } Y_{b}^*>0
\end{array} ,\right.
\end{equation}

where $Y^*_{b} \sim \mathcal{N}(\mu_b^*, 1)$. The underlying variable $Y^*_{b}$ is assumed to follow a standard normal distribution $\mathcal{N}(0, 1)$, as this avoids any loss of generality compared to using a normal distribution with an arbitrary mean and standard deviation. 

We employ a probit regression model for the conditional distribution of the binary outcome $Y_b$. The logistic model is often favored for its intuitive interpretation of coefficients in terms of odds ratios. However, we choose the probit link for its mathematical convenience, particularly the availability of closed-form solutions in certain multivariate model calculations. Despite these differences, the probit and logistic regression models generally yield similar predictions~\parencite{patnaik2013selection}.

\par 

The joint method models the joint probability by using a bivariate linear regression framework, structured as follows:

\begin{equation}
\begin{pmatrix}
Y_{c} \\
Y^*_{b}
\end{pmatrix}
\sim \mathcal{N} \left( 
\begin{pmatrix}
\mu_{c} \\
\mu^*_b
\end{pmatrix}, 
\begin{pmatrix}
\sigma_c^2 & \rho_b \sigma_c  \\
\rho_b \sigma_c & 1
\end{pmatrix}
\right),
\end{equation}

where the correlation between two outcomes is represented as $\rho_b$. The joint bivariate normal distribution of $Y^*_{b}$ and $Y_{c}$ can be expressed as the product of the marginal distribution of $Y_{c} \sim \mathcal{N}(\mu_c, \sigma_c^2)$ and the conditional distribution of $Y^*_{b}$ given $Y_{c}$:

\begin{equation}
Y^*_{b} \mid Y_{c} = y_{c} \sim \mathcal{N} \left( \mu^*_b + {\frac{\rho_b }{\sigma_c}} (Y_{c} - \mu_{c}),\ (1 - {\rho_b}^2) \right),
\end{equation}

where $\mu^*_b=\beta^*_{b0}+\beta^*_{b1}\cdot x$ and $\mu_c=\beta_{c0}+\beta_{c1}\cdot x$, representing the dose-response relationships for binary and continuous outcomes, respectively. The conditional variance needs to be standardized to ensure identifiability in our bivariate probit model. This implies

\begin{equation}
E(Y_{b} \mid Y_{c} = y_{c}) = P(Y^*_{b}>0 \mid Y_{c} = y_{c}) =\Phi \left(\frac{\mu^*_b +\frac{\rho_b}{\sigma_c}(Y_{c} - \mu_{c})}{\sqrt{(1 - {\rho_b}^2)}} \right).
\label{eq:cond_before}
\end{equation}

We reparameterize the model as follows:

\begin{equation}
E(Y_{b} \mid Y_{c} = y_{c}) = P(Y^*_{b}>0 \mid Y_{c} = y_{c}) =\Phi \left({\mu_b + {\tau}(Y_{c} - \mu_{c})} \right),
\label{eq:cond}
\end{equation}

where $\mu_b=\beta_{b0}+\beta_{b1}\cdot x$ (with $\beta_{b0}=\frac{\beta^*_{b0}}{\sqrt{(1 - {\rho_b}^2)}}$ and $\beta_{b1}=\frac{\beta^*_{b1}}{\sqrt{(1 - {\rho_b}^2)}}$),  
$\mu_c=\beta_{c0}+\beta_{c1}\cdot x$ and $\tau=\frac{\rho_b}{\sigma_c\sqrt{1-{\rho_b}^2}}$. This reparameterization addresses identifiability concerns by ensuring that the model parameters are uniquely determined~\parencite{choe2020identification}. 

In this context, $\mathbf{\beta_b}=(\beta_{b0},\ \beta_{b1})$ represents the effects of the covariates on the probit of $\mu_{b}$, conditional on $Y_{c}$. Consequently, the corresponding marginal effects, denoted as $\mathbf{\beta^*_b}=(\beta^*_{b0},\ \beta^*_{b1})$, can be derived by integrating over $Y_c$ and rescaling the conditional effects accordingly. Specifically, this rescaling yields:

\begin{equation}
\beta_{b0}^*=\beta_{b0}/\sqrt{(1 + \sigma_c^2\tau^2)}\ \text{and }\beta_{b1}^*=\beta_{b1}/\sqrt{(1 + \sigma_c^2\tau^2)}.  
\label{eq:marginal}
\end{equation}

\par

We consider the Joint9d method, which incorporates eight values of biomarkers, and the Joint2d method, which includes a single biomarker value. The Probit method, relying solely on binary toxicity outcomes, serves as the baseline for comparison. Further details about derivation of the Joint9d method are provided in Appendix~\ref{sup:joint9d}. Maximum likelihood estimation (MLE) for all methods is performed using the \textit{bbmle}\parencite{bbmle} package in R\parencite{Rversion}, employing the \textit{nlminb} optimization algorithm.

\section{Modeling challenges in dose-finding studies}\label{sec:problem}

Parameter estimation within the frequentist framework can be challenging, especially when data are limited, as is frequently the case in the early stages of dose-finding studies. In this section, we highlight key issues that arise during model fitting. To illustrate these challenges, we present two hypothetical trial realizations in phase I, each demonstrating a common estimation problem encountered in dose-finding studies.

% The initial stage remains consistent across the different methods employed in Stage 2, ensuring comparability, as demonstrated in Section~\ref{subsec:algorithm}.

For illustration, we consider a hypothetical trial with five investigated doses and assume a skeleton of
$p_0=(0.25, 0.35, 0.45, 0.55, 0.65)$ in a hypothetical trial realization, assuming equal spacing between adjacent doses. %The skeleton can be elicited from clinical expertise or historical data~\parencite{neuenschwander2015bayesian}. Alternatively, in the absence of external information, it can be determined through calibration methods~\parencite{lee2009model}. In this example, we adopt a skeleton with a fixed increment in the probability of toxicity between consecutive doses for simplicity. For a more rigorous specification of the skeleton, we refer readers to approaches that integrate expert elicitation, statistical considerations, or a combination of both~\parencite{ewings2022practical}. 
We set the target response rate to $\phi_T=0.3$. %analogous to the choice of target probability used in similar phase I dose-finding studies~\parencite{neuenschwander2008critical}. 
The trial is assumed to enroll 20 cohorts sequentially, with three patients per cohort.

%We refer to the two-parameter probit model in Table~\ref{tab:ModelDose} as the Probit method, which serves as the baseline for joint methods.

%This is also relevant for joint models, where the number of observations required to fit the model depends heavily on the underlying data pattern~\parencite{altzerinakou2021change}.

\subsection{Example Trial 1: Anchoring at a lower dose with a step function}\label{subsec:stuck}

%For illustration, we consider a scenario involving a single trial with five dose levels, where the probability of toxicity is given as $\pi = (0.05, 0.10, 0.15, 0.30, 0.45)$. 

%The target dose $d^T$ is defined as the dose level closest to $\phi_E=0.3$, which corresponds to $d_4$ in this example. We use $p_0=(0.25, 0.35, 0.45, 0.55, 0.65)$, with an initial assumption of fixed discrepancies between doses. Data generation adheres to the optimal benchmark approach \parencite{o2002non, mozgunov2020benchmark}, assuming complete information for each patient across all dose levels. For simplicity, we consider completely irrelevant ctDNA values ($\rho_b=0$) in this section and skip data generation for ctDNA. More details are provided in Section~\ref{sec:simulation}. 

\begin{figure}[H]
    \centering
    \includegraphics[scale=0.5]{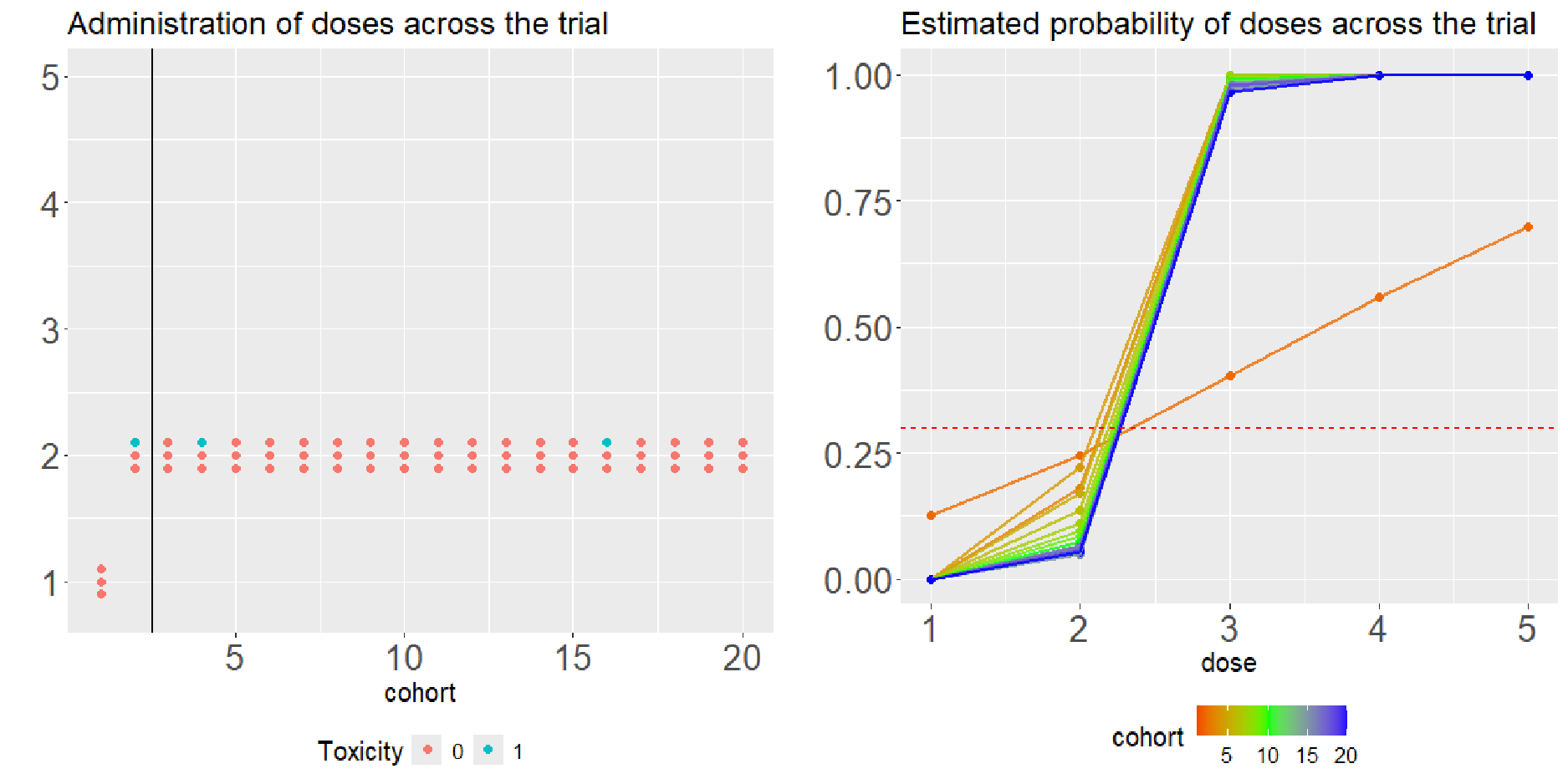}
    \caption{Illustration of a potential anchoring problem in a hypothetical trial using the Probit method. Stage 1 includes two cohorts, followed by Stage 2, which applies the two-parameter Probit model.}
    \label{fig:stuck_initial1_probit}
\end{figure}

In the example trial shown in Figure~\ref{fig:stuck_initial1_probit}, the initial stage involves two cohorts to satisfy the heterogeneity requirement: no DLTs are observed in the first cohort, while one DLT is recorded in the second. This results in a quasi-separation data configuration, as summarized in Table~\ref{tab:dataconfig}. Consequently, the two-parameter probit model repeatedly selects $d_2$ from the second cohort onward, persisting until the trial conclusion. This outcome arises because the model produces a step function, as depicted in the right panel of Figure~\ref{fig:stuck_initial1_probit}. The estimated probability of toxicity for $d_1$ and $d_2$ remains below 0.3, while the probabilities for $d_3$ through $d_5$ are estimated to approach 1. As a result, doses $d_3$ through $d_5$ are never explored during the trial.

We refer to this phenomenon as the \textit{anchoring problem} in dose-finding studies, where dose escalation becomes stalled as a result of the step function of the estimated probability for investigated doses. Once this issue arises, further dose escalation to higher levels becomes infeasible, as these doses are perceived to be too distant from the target probability $\phi_T$. Consequently, higher doses remain unexplored, and subsequent cohorts are unable to advance the trial. Ultimately, this can lead to an underdosing problem, depending on the target dose in a given scenario.

This phenomenon is more broadly known as the \textit{separation problem} in parameter estimation, where the covariate space (i.e., the investigated doses) is effectively divided into two distinct categories by the step function, leading to an infinite value (i.e., complete separation) or an extremely large value (i.e., quasi-complete separation) for the MLE of the covariate. This issue was also highlighted in the original CRML paper~\parencite{o1996continualcrml}, which notes that when higher dose levels are the correct choice, `\textit{the CRML tends to spend too much time at lower levels due to slow escalation caused by grouping}'. In this context, `grouping' refers to the partitioning or separation of doses by the step function.

Paoletti and Kramar~\parencite{paoletti2009comparison} demonstrated that the two-parameter model fails to identify the correct dose in challenging scenarios where lower doses yield only non-DLT. This issue is illustrated in greater detail by the example trial shown in Figure~\ref{fig:stuck_initial1_probit}, which provides a concrete visualization of the data conditions that lead to parameter nonidentification and consequently incorrect dose selection.

The joint modeling methods introduced in Section~\ref{subsec:Framework} have the potential to mitigate the anchoring problem under identical initial-stage conditions. With the same number of patient cohorts in the initial stage, the inclusion of biomarker information can reduce the risk of outcome-driven separation in the covariate space when fitting joint models. For illustrative purposes, we first consider an example in which the biomarker is completely irrelevant (i.e., $\rho_b = 0$). Additional details on data generation under various assumptions are provided in Section~\ref{sec:simulation}; we omit them here, as they are not central to the current discussion.

\begin{figure}[H]
    \centering
    \includegraphics[scale=0.5]{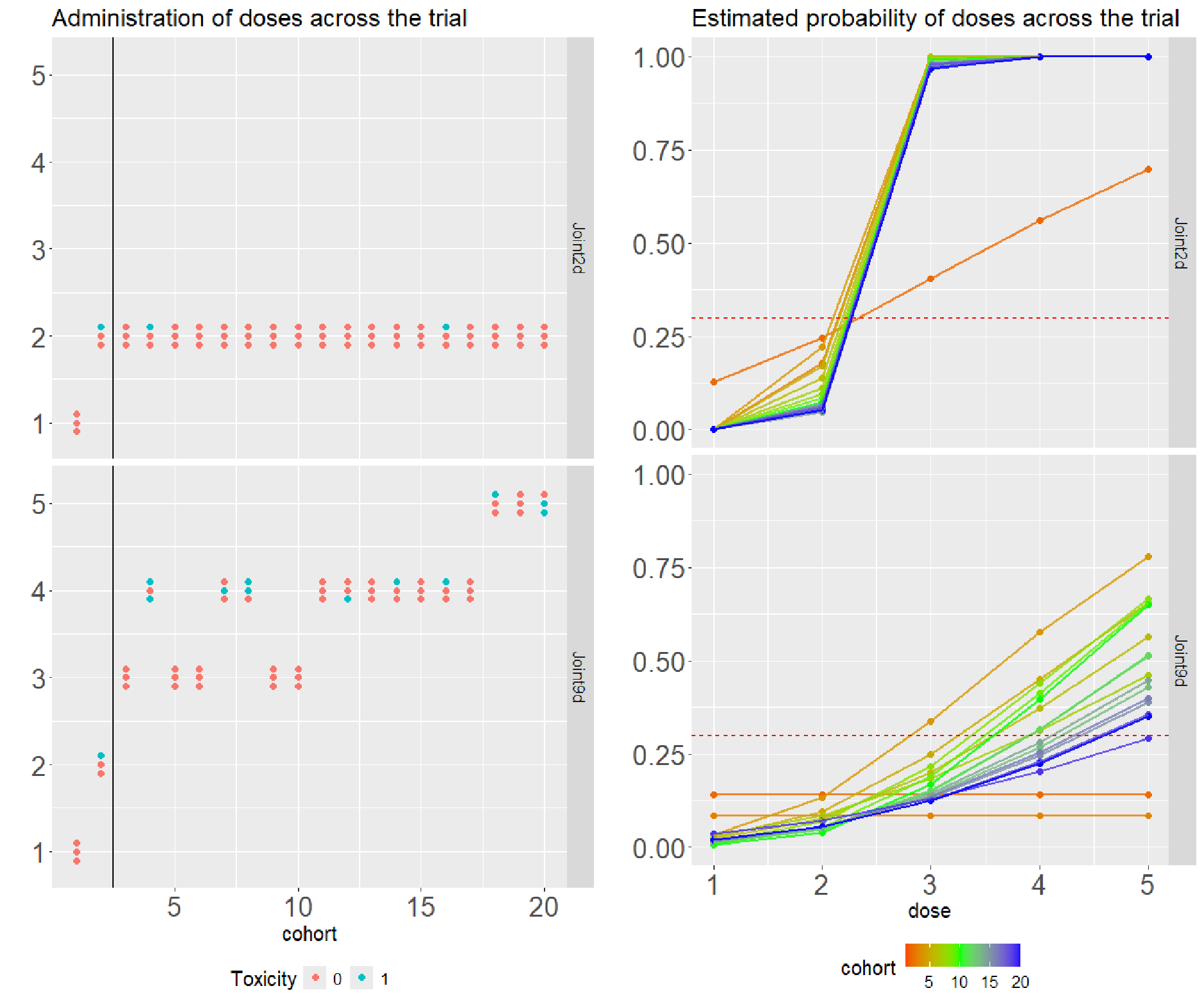}
    \caption{Illustration of the potential anchoring problem in a hypothetical trial using joint modeling approaches. Stage 1 includes two cohorts, and Stage 2 implements the Joint2d and Joint9d methods based on the two-parameter Probit model.}
    \label{fig:stuck_initial1_joint}
\end{figure}

This example is visualized in Figure~\ref{fig:stuck_initial1_joint}, which presents the same hypothetical trial under the two-stage design described in Section~\ref{subsec:algorithm}, now implementing the Joint2d and Joint9d methods in Stage 2. All other aspects of the hypothetical trial are comparable to those in Figure~\ref{fig:stuck_initial1_probit}. With the same initial stage consisting of two patient cohorts to satisfy the heterogeneity requirement, dose administration under the Joint2d method mirrors that of the Probit method in Figure~\ref{fig:stuck_initial1_probit}, as the step function in the estimated probability persists. However, the Joint9d method promotes dose escalation starting from the 3rd cohort, immediately after the initial stage, enabling a more efficient exploration of higher dose levels. As more cohorts are enrolled, the estimated probabilities are continuously updated, ultimately leading to the selection of the highest dose, $d_5$, as the recommended dose.

This trial example illustrates the estimation problem encountered with the likelihood-based approach due to the limited dataset typical of the early stages of dose-finding studies. As discussed in Section~\ref{subsec:algorithm}, considering the mutually exclusive data configurations, complete separation occurs when the predictor variables perfectly predict the outcome, leading to infinite MLEs. Quasi-complete separation arises when the outcome is perfectly predicted for a subset of the data, also resulting in problematic MLEs. In contrast, the overlap condition—where the predictor variables do not perfectly predict the outcome—ensures the existence of finite MLEs. The heterogeneity requirement, defined as the presence of both DLTs and non-DLTs, is motivated to avoid separation issues. However, it is slightly weaker than the overlap condition, as noted by O'Quigley~\parencite{o2006theoretical}.

%In this individual trial, we implemented the requirement of heterogeneity during the first stage of the trial as described in Section~\ref{subsec:algorithm}. This requirement, well-acknowledged in the literature, originates from the proposal of the CRML approach~\parencite{o1996continualcrml}, which was based on the requirement for the existence of a maximum likelihood estimate (MLE) for the chosen model.

%The issue of MLE existence in logistic regression has been extensively studied. Silvapulle~\parencite{silvapulle1981existence} and Albert and Anderson~\parencite{albert1984existence} identified a certain degree of overlap in the data as a necessary and sufficient condition for the MLE to exist. Similarly, Lesaffre and Kaufmann~\parencite{lesaffre1992existence} proposed a corresponding condition for the probit model, which aligns with that of Albert and Anderson~\parencite{albert1984existence} for the logistic model. This condition can be formulated in terms of the separation of observation points within the covariate space. Essentially, a certain degree of \textit{intermixed} or \textit{overlapping} data points is required to ensure the existence of the MLE in both logistic and probit models, a concept known as the \textit{separation condition}.

In dose-finding studies, limited datasets often face separation or quasi-separation issues, even when the heterogeneity requirement is satisfied. For example, lower doses may result in only non-DLTs, while DLTs are observed only at the highest doses. This data configuration can lead to inappropriate estimates of the slope parameter and, consequently, a step function in the estimated probabilities, which can hinder dose-escalation decisions. Figure~\ref{fig:stuck_initial1_probit} and Figure~\ref{fig:stuck_initial1_joint} illustrate these challenges using the Probit and Joint2d methods. Incorporation of higher-dimensional biomarker values (e.g., Joint9d in Figure~\ref{fig:stuck_initial1_joint}) offers a practical solution in this specific example trial, as higher-dimensional biomarker values, although noisy, help prevent the covariate space (including dose and continuous outcomess, as defined in the conditional probit model in Equation~\ref{eq:Joint9d_cond}) from becoming separated by the outcomes.

\subsection{Example Trial 2: Equal estimated probability problem}\label{subsec:eq}

Another challenge in parameter estimation arises when the dose-toxicity curve appears flat rather than monotonic after observing outcomes. This can happen if the empirical toxicity probabilities (i.e., observed frequencies) are similar across doses or if the data inadvertently suggest a decreasing dose-toxicity trend.

\begin{figure}[H]
    \centering
    \includegraphics[scale=0.5]{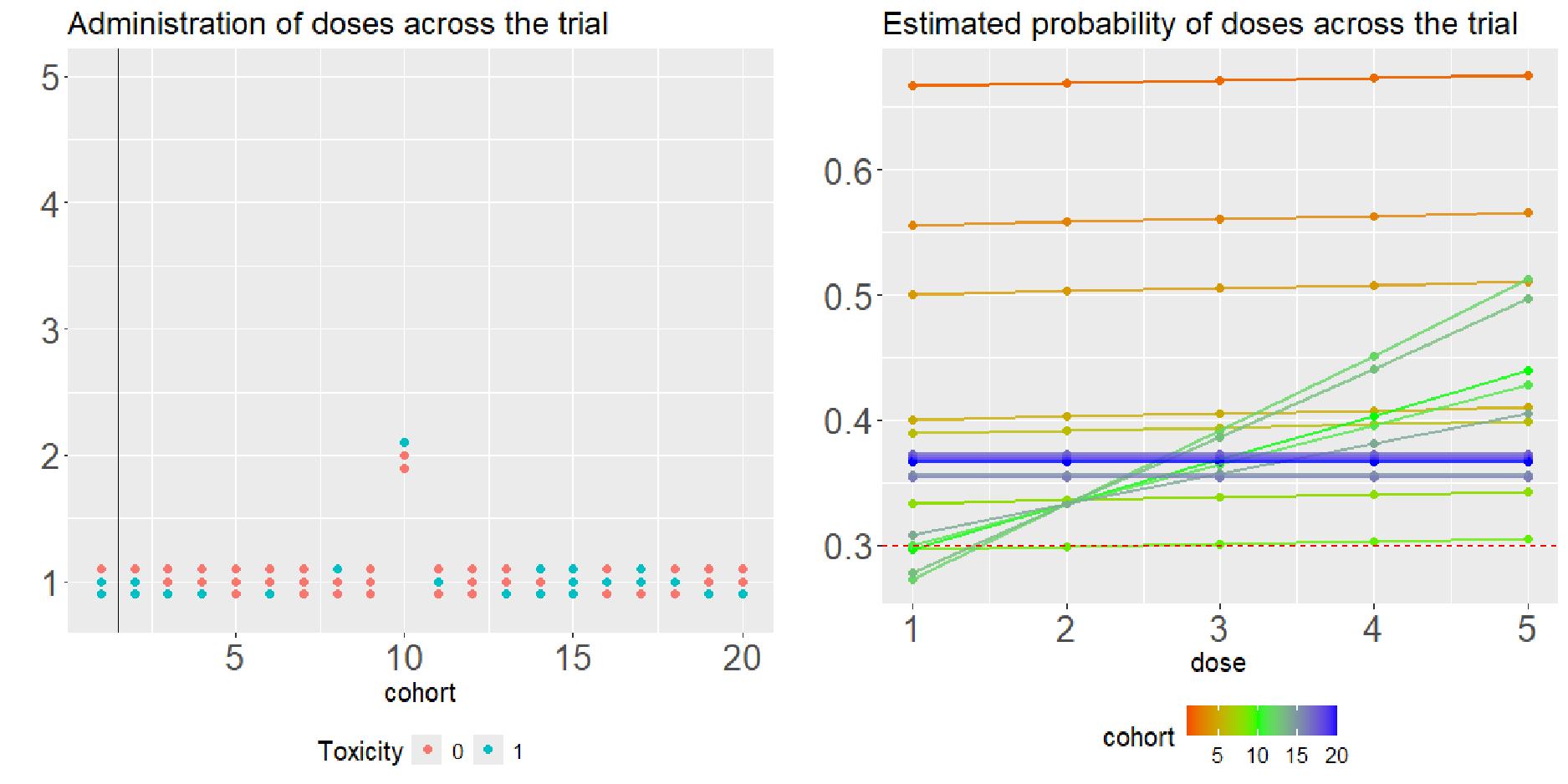}
    \caption{Illustration of the potential issue of equal estimated probabilities in a hypothetical trial using the Probit method. Stage 1 includes one cohort, followed by Stage 2, which applies the two-parameter Probit model.}
    \label{fig:example_eq_probit}
\end{figure}

Figure~\ref{fig:example_eq_probit} illustrates a hypothetical trial following the two-stage design described in Section~\ref{subsec:algorithm}, where Stage 1 consists of a single cohort. As noted in Section~\ref{sec:methods}, the non-negativity constraint on $\exp(\beta_{b1})$ results in estimated toxicity probabilities that remain similar across all dose levels throughout the trial. This constraint tends to arise in extreme scenarios where the doses have nearly indistinguishable toxicity profiles—either uniformly very safe or uniformly overly toxic. In such cases, the impact on target dose selection is generally minimal, since the recommended dose may still be appropriate despite the lack of precision in toxicity estimates.

\begin{figure}[H]
    \centering
    \includegraphics[scale=0.5]{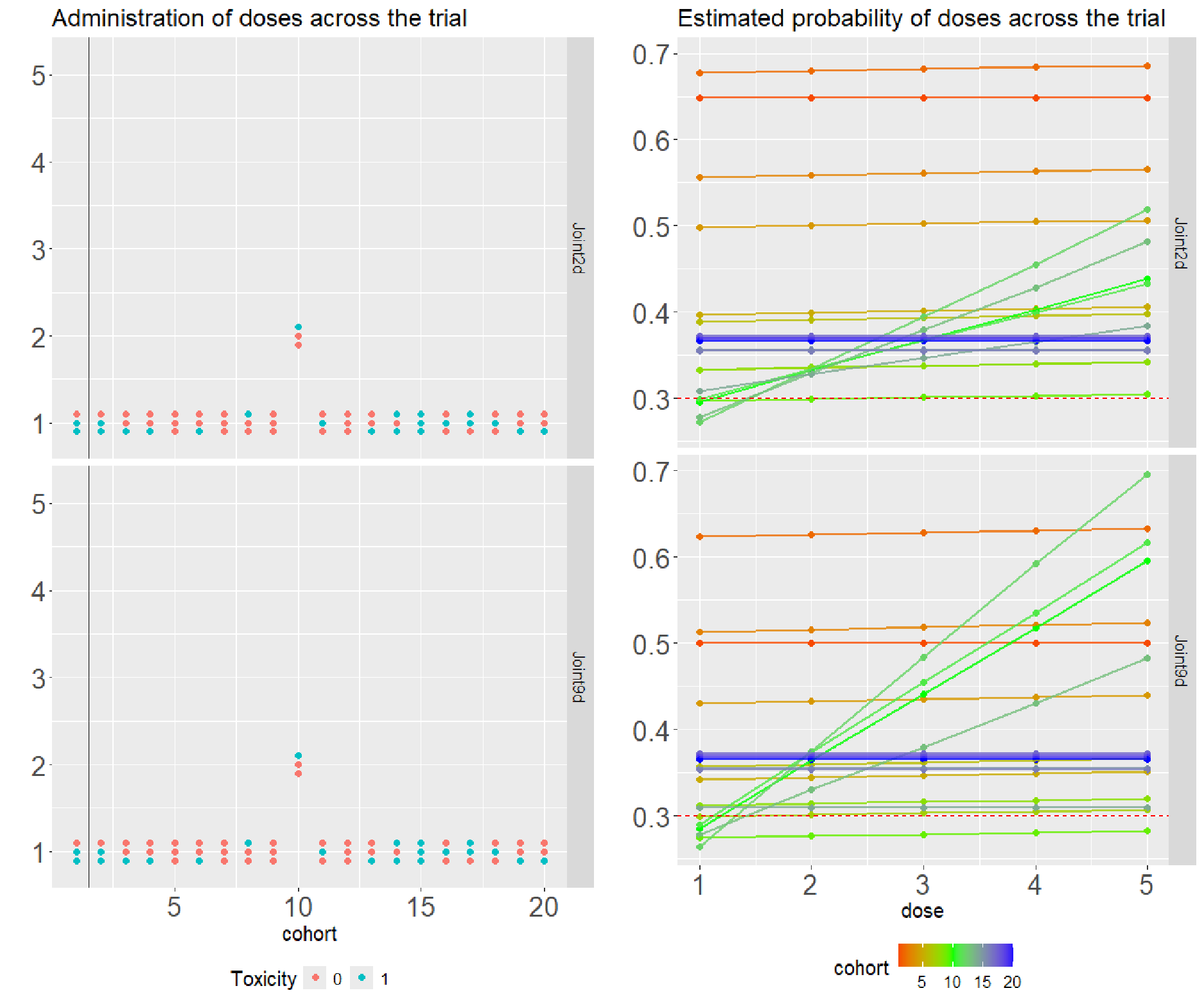}
    \caption{Illustration of the potential equal estimated problem in a hypothetical trial using the joint methods.  Stage 1 includes one cohort and Stage 2 applies the Joint2d and Joint9d methods based on the two-parameter probit model.}
    \label{fig:example_eq_probit_joint}
\end{figure}

Similarly, Figure~\ref{fig:example_eq_probit_joint} demonstrates that adding irrelevant biomarker variable (i.e., $\rho_b=0$) does not enable the Joint2d and Joint9d method to differentiate between doses when the frequencies of the observed toxicity outcomes are similar. The dose-toxicity curve eventually flattens again in the end.

%\begin{table}[!ht]
%\centering
%\begin{tabular}{c c c c c c}
%\hline
% $d$ & $d_1$  & $d_2$ & $d_3$ & $d_4$ & $d_5$ \\ 
%\hline
%$Y_b=0$ & 2 & 3 & 2 & 3 & 38 \\ 
%$Y_b=1$ & 1 & 0 & 1 & 0 & 10\\ 
%\hline
%\end{tabular}
%\caption{Dose and responses data of an example of the problem of equal estimated probability.}
%\label{tab:example_eq}
%\end{table}

This example highlights the issue of equal estimated probabilities across doses, yet different methods ultimately manage to capture information from data and lead to the desired dose recommendation by the end of the trial. All methods attempt to depict the monotonic curve in the middle of the trial after approximately 10 cohorts. However, in the end, it reverts to a flat line, overwhelmed by the information transformed from the binary outcomes. 

\par

Overall, there are two primary types of estimation challenges in dose-finding studies, where the final dose-toxicity curve is either estimated as a step function or a flat line. A key consideration is whether the joint methods with additional biomarker values can consistently resolve estimation challenges across different scenarios and how its performance is affected when biomarker values contain meaningful information rather than just noise. Moreover, while the original proposal claiming that overall fit was not the primary objective of phase I studies~\parencite{o1990continual}, it is crucial to emphasize that these estimation problems, especially the separation problem, should ideally be avoided from the outset to ensure a fair comparison across different methods. 

We intend to address these two issues in this section in a different way. It is essential to avoid step functions, as they can hinder the dose-escalation process. While a flat curve is unexpected, it can still provide valuable insights based on empirical toxicity outcomes. Thus, we focus on remedies for the anchoring problem, while still illustrating the proportion of equal probability estimation cases in comprehensive simulations.

\section{Remedies for modeling challenges}\label{sec:remedies}

To address the parameter estimation challenges with limited data in Section~\ref{subsec:stuck}, two general approaches can be considered. One solution is to increase the dataset size during the initial stage, which introduces greater complexity but may ensure the existence of the MLE in the modeling stage. Alternatively, the model may be simplified, such as by adopting a one-parameter probit model or a empiric model. For binary outcome analysis with small sample sizes, established techniques such as Firth’s logistic regression~\parencite{firth1993bias} and Jeffreys' prior~\parencite{jeffreys1946invariant} have been developed. 

In this study, we focus specifically on addressing the estimation problem in the Probit method and its corresponding joint methods within the frequentist framework. In addition to increasing the complexity of the initial stage (Section~\ref{subsec:problem-initial}) and employing a one-parameter model (Section~\ref{subsec:reme-1para}), we explore the calibration of the dose labels (Section~\ref{subsec:cali-label}).

\subsection{Initial stage}\label{subsec:problem-initial}

It is highly unlikely that the initial stage in dose-finding designs will satisfy the overlap requirement, particularly in cases where lower doses consistently result in non-DLTs. To address this, we increase the complexity of the initial stage by incorporating more occurrences. For example, a rule-based `2+2' initial stage, which requires at least two DLTs and two non-DLTs, has been used by Altzerinakou and Paoletti~\parencite{altzerinakou2021change}. We refer to this more complex initial stage as Initial2, while the simpler heterogeneity requirement~\parencite{o1996continualcrml} is referred to as Initial1. Similarly, more complex stages such as Initial3 (at least three DLTs and three non-DLTs) and Initial4 (at least four DLTs and four non-DLTs) can also be implemented in Stage 1.

\subsection{One-parameter model}
\label{subsec:reme-1para}

The one-parameter models, including the empiric model and the one-parameter probit model with a fixed intercept (see Table~\ref{tab:ModelDose}) are commonly used in the CRM literature. The empiric model is widely recognized for its superior performance compared to the two-parameter logistic model (and similarly, the two-parameter probit model)~\parencite{paoletti2009comparison}. A simulation study by Cheveret~\parencite{chevret1993continual} illustrates that the performance of CRM via power model could be improved by the use of the one-parameter logistic model, although results were rather sensitive to the choice of the fixed parameter $a_0$. 

However, one often-overlooked aspect in previous investigations is that the intercept $a_0$ implicitly encodes information about a \textit{reference dose} $x^*=0$. The probability of toxicity at this reference dose is determined at $\pi(x^*)=\psi(x^*)=\psi(a_0)$, independent of the parameter estimation for $\beta_{b1}$. As a result, the dose-toxicity curve and the estimation of toxicity at investigated doses can be significantly influenced by the reference dose $x^*$, equally, the intercept $a_0$. When $x^*$ falls within the range of dose labels, the fixed probability ${\pi}(x^*)$, determined by $a_0$, acts as a dividing point in the dose-response relationship. For example, if $a_0=0$, the skeleton $p_0=(0.25, 0.35, 0.45, 0.55, 0.65)$ corresponds to the transformed dose labels $x=(-0.67, -0.39, -0.10, 0.15, 0.39)$, with the division occurring between $d_3$ and $d_4$. Specifically, $\hat{\pi}(x_j) < 0.5$ holds for all doses below $d_4$. More generally, the dichotomization of dose labels and the estimated probability ${\pi}(x^*)$ depends on the chosen model or the value of $a_0$, as shown in Equation~\ref{eq:dicotomization}.

\begin{equation}
    \hat{\pi}(x_j) 
\begin{cases}
    < {\pi}(x^*) &  x_j < x^*\ (x^*=0) \\
    > {\pi}(x^*) & x_j > x^*\ (x^*=0).
\end{cases}
\label{eq:dicotomization}
\end{equation}

\begin{figure}[H]
    \centering
    \includegraphics[scale=0.4]{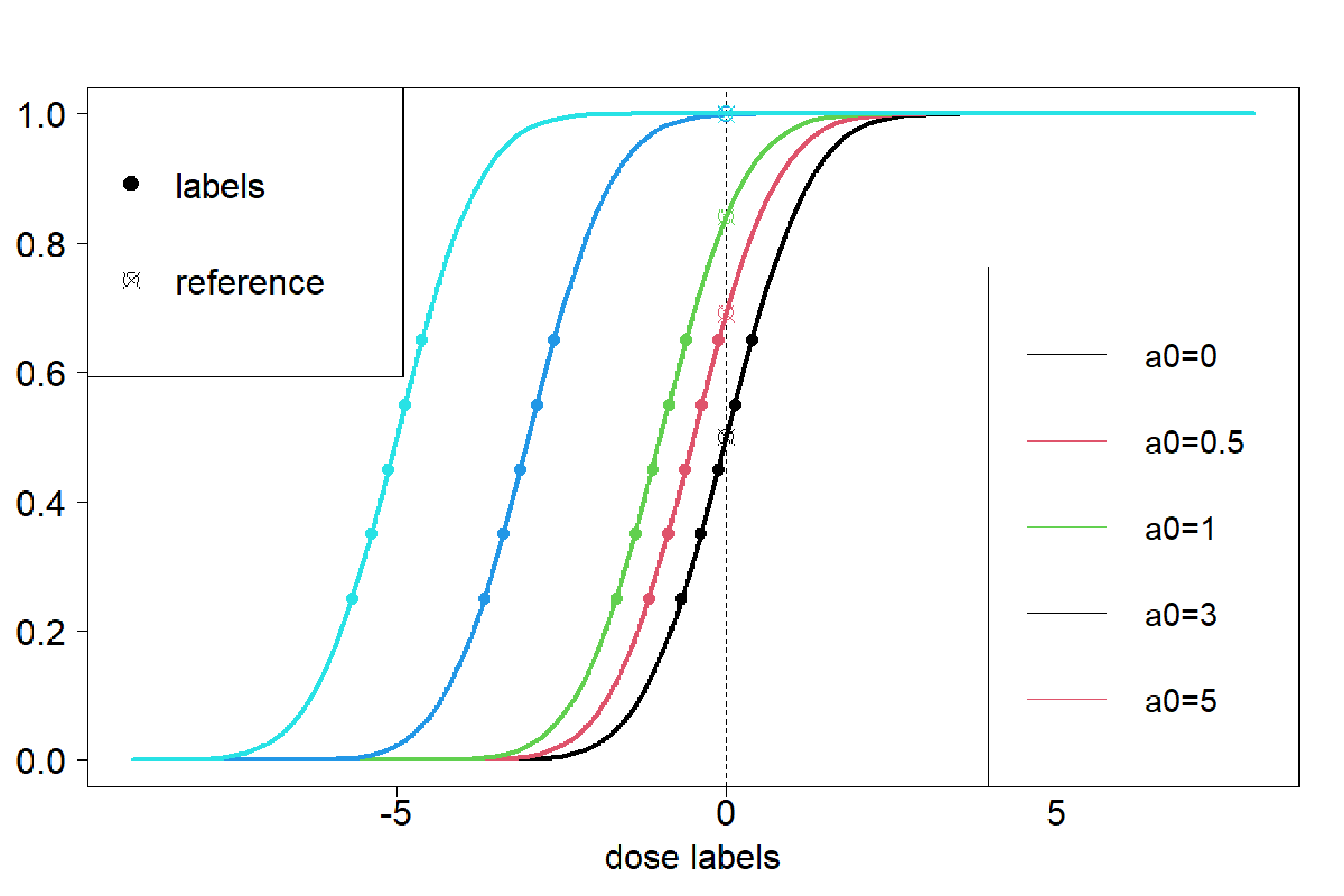}
    \caption{The one-parameter probit model with various values of $a_0$. Corresponding dose labels and the reference dose are illustrated. Reference dose with ${\pi}(x^*)=(0.5, 0.691, 0.841, 0.998, 1)$ corresponding to $a_0=(0, 0.5, 1,3,5)$.}
    \label{fig:One_Para}
\end{figure}

Figure~\ref{fig:One_Para} demonstrated the one-parameter probit model under different choices of $a_0$, each leading to distinct sets of transformed dose labels (see Table~\ref{tab:ModelDose}). Among the choices shown, $a_0=0$ is the only one that results in the dichotomization of dose levels. In the case $a_0=0.5$, the transformed dose labels, $x = (-1.17, -0.89, -0.63, -0.37, -0.11)$, are not dichotomized by $x^*=0$. Instead, the reference dose $x^*$ acts as an upper limit, with $\pi(x^*) = 0.69$. In such cases, the model may lead to underestimation if any dose $x_j$ satisfies $\pi(x_j) > 0.69$.

Ultimately, we conclude that a sufficiently large value of $a_0$ is essential for the one-parameter probit model to prevent the reference dose $(x^*, \pi(x^*))$ from dichotomizing dose levels or imposing inappropriate upper limits on the estimated probabilities of response. However, selecting an excessively large value for $a_0$ is not always advisable, as it may result in excessively large arguments in the normal cumulative distribution function of the probit model or logistic model, potentially causing the estimated probabilities for all doses to converge to one~\parencite{demidenko2001computational}. This issue reminds us that choosing an overly large $a_0$ does not necessarily resolve convergence issues and may, in fact, exacerbate them. Chevret~\parencite{chevret1993continual} investigated the choice of $a_0$ in the one-parameter logistic model within a Bayesian framework and recommended performing simulation studies to assess sensitivity before finalizing its value.

%This conclusion is further supported by extensive simulations presented in Section~\ref{sec:one-paraProbit}. 

%Demidenko~\parencite{demidenko2001computational} identified two primary reasons for convergence failure during the MLE procedure for probit model. The first is the non-existence of an MLE, and the second is that the argument of the normal cumulative distribution function becomes excessively large in absolute value, often due to rare or frequent event data. 

\par

\subsection{Calibration for dose labels}
\label{subsec:cali-label}

Paoletti and Kramar~\parencite{paoletti2009comparison} evaluated the impact of the coding system, which codes value for dose level, on the operating characteristics of the empiric model. Their findings indicated that the choice of coding system can substantially influence the performance of the CRM when using the empirical model.

A systematic calibration of the implemented dose levels—incorporating both the skeleton and the original model—may further improve the operating characteristics of the Probit method, as well as the joint methods. Future comparisons of these methods could also be expanded to include dose label calibration.

%By considering the dose increment between two adjacent levels in a dose sequence, the working model can be adapted to some extent~\parencite{paoletti2009comparison}. 

We perform calibration of dose labels based on both the model parameters and the skeleton of doses, and then select the set of dose labels that optimizes the geometric mean of the PCS results across the investigated scenarios, using the calibrated skeleton and model parameters. The grid search for the calibrated skeleton and model parameters is carried out through an iterative process, starting with a broad range of values and progressively refining them to identify the optimal combination. For each combination of skeleton and model parameters, the proportion of correct selections (PCS) is calculated for each scenario presented in Table~\ref{tab:scenarios}. The geometric mean of PCS across all investigated scenarios is then computed. The combination that achieves the highest PCS is selected, and the corresponding dose labels are determined based on the chosen method. The grid values can be informed by prior experience, if available.

\section{Simulation}\label{sec:simulation}

To investigate the contribution of additional biomarker values in dose-finding studies, we evaluate the performance of the joint methods—specifically, the Joint9d and Joint2d approaches—through a comprehensive simulation study, using the Probit method as a baseline for comparison. The data generation process and simulation settings are detailed in Section~\ref{subsec:ctDNA} and Section~\ref{subsec:scenarios}. The operating characteristics used to assess the performance of the methods are introduced in Section~\ref{subsec:metrics}. Furthermore, to evaluate the effectiveness of the remedies proposed in Section~\ref{sec:remedies}, we first applied all remedies within the same hypothetical trial introduced in Section~\ref{subsec:stuck} and presented the corresponding simulation results across various scenarios. 

\par
\subsection{Data generation}\label{subsec:ctDNA}

We consider a study with five doses, $d_1< d_2< d_3< d_4 < d_5$, corresponding to dose labels $x_1< x_2< x_3< x_4 < x_5$. The transformation of these dose labels is outlined in Table~\ref{tab:ModelDose} and is applied uniformly to both the Probit and joint methods. The trial design includes eight week evaluation window for binary toxicity outcomes and biomarker measurements are taken weekly, starting from week 1 (recruitment) and continuing until the toxicity evaluation at week 8. Consequently, eight biomarker measurements are collected by the time the binary outcomes are assessed. Dose administration decisions for the subsequent cohort are made once the toxicity data becomes available.

The data generation process uses complete information about patient profiles as the optimal benchmark approach, which was originated in O’Quigley et al.~\parencite{o2002non} and extended in Mozgunov et.al.~\parencite{mozgunov2020benchmark}. Assuming that the outcomes of each patient can be observed at all dose levels, it allows to minimize simulation error and provide a more accurate comparison between methods. 

Specifically, we generate correlated random variables from a standard multivariate normal distribution to represent patients' profiles, which serve as the basis for generating complete outcomes across all doses. For each patient $i$ administered dose $d_j$, a 9-dimensional random vector is drawn from a multivariate normal distribution, as specified in Equation~\ref{eq:completeInformation}. This vector incorporates correlation through the variance-covariance matrix, $\Sigma$. The patients' profiles are obtained by transforming these random variables using the standard normal probability density function. Subsequently, complete responses are derived via quantile transformations using the corresponding cumulative distribution functions; that is, $Y_{ci,t}=\Phi^{-1}(u_{ci,t})$ for the countious outcomes. Similarly for binary outcomes, $Y_{bi}=1-F^{-1}(u_{bi})$, where $F$ is CDF of Bernoulli distribution. In this context, the parameter $\rho_c$ and $\rho_b$ correspond to the relationship in the R.V. in Equation~\ref{eq:completeInformation}. Specifically, $\rho_c$ represents the autoregressive coefficient under the AR(1) assumption, while $\rho_b$ quantifies the correlation between $x_{bi}$ and $x_{ci,8}$ corresponding to the lastly evaluated biomarker value. To demonstrate how we generate complete patient information, we present an example involving two hypothetical patients with $Y_{ci,8}$ and $Y_{bi}$ for patient $i=1$ and patient $i=2$. 

\begin{equation}
\begin{pmatrix}
u_{ci,1} \\
u_{ci,2} \\
u_{ci,3} \\
u_{ci,4} \\
u_{ci,5} \\
u_{ci,6} \\
u_{ci,7} \\
u_{ci,8} \\
u_{bi}
\end{pmatrix}
\sim \mathcal{N} \left( 
\begin{pmatrix}
0 \\
0 \\
0 \\
0 \\
0 \\
0 \\
0 \\
0 \\
0
\end{pmatrix}, 
\begin{pmatrix}
1 & \rho_c & \rho_c^2 & \rho_c^3 & \rho_c^4 & \rho_c^5 & \rho_c^6 & \rho_c^7 & \rho_b^8 \\
\rho_c & 1 & \rho_c & \rho_c^2 & \rho_c^3 & \rho_c^4 & \rho_c^5 & \rho_c^6 & \rho_b^7 \\
\rho_c^2 & \rho_c & 1 & \rho_c & \rho_c^2 & \rho_c^3 & \rho_c^4 & \rho_c^5 & \rho_b^6 \\
\rho_c^3 & \rho_c^2 & \rho_c & 1 & \rho_c & \rho_c^2 & \rho_c^3 & \rho_c^4 & \rho_b^5 \\
\rho_c^4 & \rho_c^3 & \rho_c^2 & \rho_c & 1 & \rho_c & \rho_c^2 & \rho_c^3 & \rho_b^4 \\
\rho_c^5 & \rho_c^4 & \rho_c^3 & \rho_c^2 & \rho_c & 1 & \rho_c & \rho_c^2 & \rho_b^3 \\
\rho_c^6 & \rho_c^5 & \rho_c^4 & \rho_c^3 & \rho_c^2 & \rho_c & 1 & \rho_c & \rho_b^2 \\
\rho_c^7 & \rho_c^6 & \rho_c^5 & \rho_c^4 & \rho_c^3 & \rho_c^2 & \rho_c & 1 & \rho_b \\
\rho_b^8 & \rho_b^7 & \rho_b^6 & \rho_b^5 & \rho_b^4 & \rho_b^3 & \rho_b^2 & \rho_b & 1
\end{pmatrix}
\right)
\label{eq:completeInformation}
\end{equation}

\textit{Example:} Consider an extreme case where $\rho_b$ indicates a perfect relationship between $u_{ci,8}$ and the toxicity outcomes $u_{bi}$. The bivariate normal vector, with mean $\mu_i=(0,0)$ and covariance matrix $\Sigma=\begin{pmatrix}
  1 & \rho_b\\ 
  \rho_b & 1
\end{pmatrix}$ is initially generated as $(x_{bi},\ x_{ci,8})$. In this example (see Table~\ref{tab:complete_profile}), the first simulated patient $i=1$ has a toxicity profile of $u_{bi} = \Phi(0.46)=0.68$ and an biomarker profile $u_{ci,8} = \Phi(0.46)=0.68$. These values correspond to the toxicity response $Y_{bi}=1-F^{-1}(u_{bi},\  \pi_T)$ and biomarker values 
 $Y_{ci,8}=\Phi^{-1}(u_{ci,8},\ \mu_{c8},\ \sigma_c=1)$. For illustrative purposes, we assume $\pi_T=(0.30,\ 0.45,\ 0.55,\ 0.65,\ 0.75)$ and $\mu_{c8}=(10,\ 8,\ 6,\ 4,\ 2)$. In the end, treatment is more toxic for the patient $i=1$ than for the patient $i=2$, indicated by a higher probability of favorable toxic responses ($Y_b=1$ in $d=5$) and generally lower biomarker values (see Table~\ref{tab:complete_profile}). It is important to note that, for illustrative purposes in the data generation process, we assume $\rho_b = 1$, indicating a perfect correlation between the profiles for binary and continuous outcomes. However, this assumption is unlikely to hold in practical applications. 

\begin{table}[H]
\centering
\begin{tabular}[8pt]{ccccc}
\hline
patient $i$ & ($x_{bi},\ x_{ci,8}$) &  ($u_{bi},\ u_{ci,8}$) &  $Y_{bi}$  &  $Y_{ci,8}$ \\
\hline
$i=1$ & (0.46, 0.46) & (0.68, 0.68) & (0,0,0,0,1) & (10.47,  8.47,  6.47,  4.47,  2.47)\\
$i=2$ & (1.34, 1.34) & (0.91, 0.91) & (0,0,0,0,0) & (11.34,  9.34,  7.34,  5.34,  3.34) \\
\hline
\end{tabular}
\caption{Example of two patients with complete information generation.}
\label{tab:complete_profile}
\end{table}

The Probit method  exclusively uses binary outcomes, $Y_b$, while the Joint2d method, in addition to $Y_b$, incorporates a single biomarker measurement, $Y_{c8}$, collected at week 8 concurrently with the binary outcomes. The Joint9d method extends this by utilizing all biomarker measurements from week 1 to week 8, along with $Y_b$.

Weekly values of biomarker are modeled as normally distributed $\mathbf{Y_c} \sim \mathcal{N}(\mathbf{\mu_{c}}, \mathbf{\Sigma_c})$, where the vector $\mathbf{\mu_{c}}$ represents the mean values $\mu_{ct}(x)$ ($t \in [1, 8]$) and $\mathbf{\Sigma_c}$ denotes the variance-covariance matrix of biomarker values. The linear relationship between time $t$, dose $x$, and $\mu_{ct}$ is assumed as 

\begin{equation}
\mu_{ct}(x)=\beta_{c0} +\beta_{c1}\cdot x+a_t\cdot t.
\label{eq:mu_ctDNA}
\end{equation}

In Equation~\eqref{eq:mu_ctDNA}, $x$ represents the candidate dose, and $\beta_{c1}$ captures the relationship between doses and mean value of biomarker responses. We assume a fixed linear relationship for patient responses in biomarker over time, $a_t$, reflecting the treatment's effectiveness in reducing biomarker values on average. A negative value of $a_t$ indicates the benefit from treatment with time. The variance-covariance matrix of the multivariate distribution, $\mathbf{\Sigma_c}$, is modeled under the assumption of autocorrelation as detailed in Section~\ref{subsec:notationctDNA}.

While this study assumes fixed linear trajectories for mean values, within-subject variance, and intra-subject variance, alternative configurations could be explored in future research. This work represents an initial investigation, highlighting the potential challenges and advantages of integrating biomarker data into dose-finding decisions within the framework of joint models.

\par
\subsection{Simulation scenarios and parameter choices}\label{subsec:scenarios}

We analyze the simulation results under varying correlations of $\rho_b$ and $\rho_c$ (see Table~\ref{tab:design_para}) across different dose-response scenario combinations presented in Table~\ref{tab:scenarios}. In this initial investigation, the variance of biomarker values is fixed at $\sigma_c = 1$.

To ensure a positive conditional variance in the Joint2d and Joint9d methods (Section~\ref{subsec:Framework}) and maintain a positive-definite covariance matrix for data generation, parameter selection is restricted. Specifically, a combination of a large $\rho_b$ and a small $\rho_c$ is not suitable. More details about the constraints on the choices of $\rho_b$ and $\rho_c$ are provided in Appendix~\ref{sup:cond_var}.

Since our primary interest lies in the impact of the correlation between biomarker and binary toxicity outcomes, we fix $\rho_c$ at a moderate value $\rho_c=0.4$ and explore simulation results for $\rho_b = (0, 0.4, 0.8)$. While this study focuses on these specific configurations, alternative settings could be explored in future research. Table~\ref{tab:design_para} also summarizes other key parameters used in the study design.

\begin{table}[H]
\small
\begin{subtable}[h]{0.55\textwidth}
\centering
{\begin{tabular}[8pt]{ccc}
\hline
Notation &  Interpretation &  Value \\
\hline
s & Number of simulations & 1000 \\
N & Total number of patients & 60 \\
n & Number of cohorts & 20 \\
m & Cohort size & 3 \\
J & Number of dose levels & 5 \\
$\rho_b$ & Correlation between $x_b$ and $x_{c,8}$ & $$(0,\ 0.4,\ 0.8)$$ \\
$\rho_c$ & Correlation between adjacent biomarker  & 0.4 \\
$\sigma_c$ & Variance of biomarker values  & 1 \\
$\beta_{c0}$ & Intercept for $\mu_{ct}$ at baseline  &  20 \\
$\beta_{c1}$ & Impact of dose on $\mu_{ct}$  &  -2 \\
$a_t$ & Impact of time on $\mu_{ct}$  &  -1 \\
\hline
\end{tabular}}
\caption{Summary of design parameters}
\label{tab:design_para}
\end{subtable}
\hfill
\begin{subtable}[h]{0.35\textwidth}
\centering
\begin{tabular}{cc}
\hline
& $\pi_T(x_j)$ \\
\hline
$S_1$ & (\textbf{0.30}, 0.45, 0.55, 0.65, 0.75)  \\
$S_2$ & (0.15 \textbf{0.30} 0.45 0.55 0.65)  \\
$S_3$ & (0.10 0.15 \textbf{0.30} 0.45 0.55)  \\
$S_4$ & (0.05 0.10 0.15 \textbf{0.30} 0.45)  \\
$S_5$ & (0.05 0.08 0.10 0.15 \textbf{0.30})  \\
\hline
\end{tabular}
\caption{Scenarios with the targeted dose}
\label{tab:scenarios}
\end{subtable}
\caption{Simulation scenarios and corresponding design parameters}
%\label{tab:scenarios}
\end{table}

\par

\subsection{Operating characteristics}\label{subsec:metrics}

For performance evaluation, we focus primarily on four metrics. First, we assess (1) \textit{the proportion of correct selections (PCS)}. In addition, to investigate the estimation issues discussed in Section~\ref{sec:problem}, we include two supplementary metrics: (2) \textit{the proportion of separation}, defined as the proportion of cases in which the estimated probabilities satisfy either $|\hat\pi(d_i) - 0| \leq 10^{-4}$ or $|\hat\pi(d_i) - 1| \leq 10^{-4}$; and (3) \textit{the proportion of equal estimated probabilities}, defined as the proportion of cases in which $|\hat\pi(d_5) - \hat\pi(d_1)| \leq 10^{-4}$. Finally, for safety considerations, we report (4) \textit{the proportion of early stopping} due to excessive toxicity.
%\subsection{Simulation results}\label{subsec:results}

%We begin by presenting the simulation results across different methods. Following this, we evaluate the performance of these methods after applying the remedies proposed in Section~\ref{sec:remedies}.

\subsection{Impact of initial stage}\label{sec:initial}

Given access to the same complete information described in Section~\ref{subsec:stuck}, we ensure a fair comparison by examining the trial in Figure~\ref{fig:stuck_initial1_probit} with identical setups but varying initial stages. In the example presented in Figure~\ref{fig:stuck_initial1_probit}, the Initial1 stage comprises two cohorts ($Y_b=(0, 0, 0, 0, 0, 1)$ and $d_j=(1, 1, 1, 2, 2, 2)$). In contrast, the Initial3 stage extends beyond the second cohort, continuing until two additional toxicities are observed. Ultimately, the Initial3 stage involves seven cohorts: one toxicity occurs in the 2nd cohort at dose level $d_2$, one, occurs in the 4th cohort at $d_3$, and the last one occurs in the 7th cohort at $d_5$ ($Y_b=(0, 0, 0, 0, 0, 1, 0, 0, 0, 1, 0, 1, 0, 0, 0, 0, 0, 0, 0, 1, 0)$ and $d_j=(1, 1, 1, 2, 2, 2, 2, 2, 2, 3, 3, 3, 3, 3, 3, 4, 4, 4, 5, 5, 5)$). 

\begin{figure}[H]
    \centering
    \includegraphics[scale=0.5]{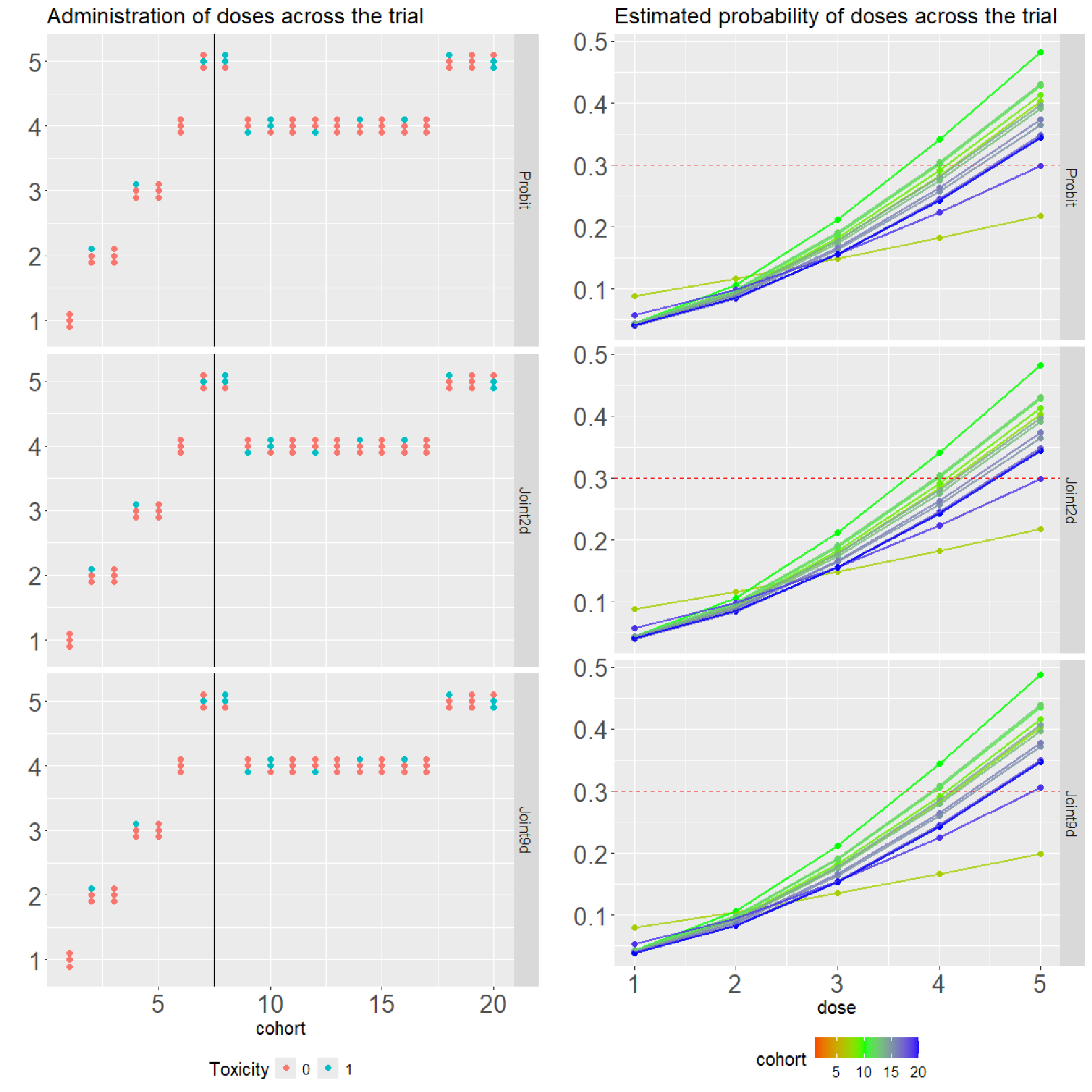}
    \caption{Illustration of a hypothetical trial without the anchoring problem following the implementation of an alternative initial stage (Initial3). Stage 1 includes seven cohorts, and Stage 2 applies methods based on the two-parameter probit model.}
    \label{fig:nonstuck_initial3_all3}
\end{figure}

Figure~\ref{fig:nonstuck_initial3_all3} illustrates how the choice of the initial stage influences the results of the dose-finding design. Every aspect other than the choice of initial stage of the hypothetical trial is comparable to that in Section~\ref{subsec:stuck}. With the Initial3, both the Probit and Joint2d methods successfully avoid the anchoring problem in Figure~\ref{fig:stuck_initial1_probit} and Figure~\ref{fig:stuck_initial1_joint}, progressing to $d_5$ by the end of the trial. The Joint9d method behaves similarly to its performance in Figure~\ref{fig:stuck_initial1_joint}, where there is no estimation problem with both Initial1 and Initial3. This example also suggests that the Joint9d method exhibits robustness to the complexity of the initial stage, while the anchoring problem of the Probit and Joint2d methods can be avoided by a more complicated initial stage. Similarly, this estimation problem can be alleviated under Inital2 as shown in Appendix~\ref{sup:Initial2}.

%In Section~\ref{subsec:problem-initial}, we demonstrated that a more complex initial stage can effectively mitigate the stuck problem in an individual trial. Here, we present simulation results that evaluate the performance of more different initial stages under general simulation settings.

To thoroughly assess the impact of the initial stage, we evaluate the PCS of the Probit method across various scenarios using different initial stages, as introduced in Section~\ref{subsec:problem-initial}. For comparison, we also include simulation results for the Empiric method, as previously investigated in related studies~\parencite{o1996continualcrml, altzerinakou2021change}. Corresponding dose-finding algorithm is the same as described in Section~\ref{subsec:algorithm}. Unlike the Probit and Logistic models, which may fail to generate reasonable estimates under specific data configurations, the Empiric model exhibits greater robustness, consistently producing reliable estimates across different initial stages. A comparison of model choices by Paoletti and Kramar~\parencite{paoletti2009comparison} concludes that a one-parameter empiric (power) model outperforms a two-parameter logistic model. This superiority becomes more pronounced as sample sizes increase, reinforcing the finding that the separation problem in the two-parameter probit or logistic models emerges early in the trial and persists, as discussed in Section~\ref{subsec:stuck}. 

\begin{figure}[H]
    \centering
    \includegraphics[scale=0.6]{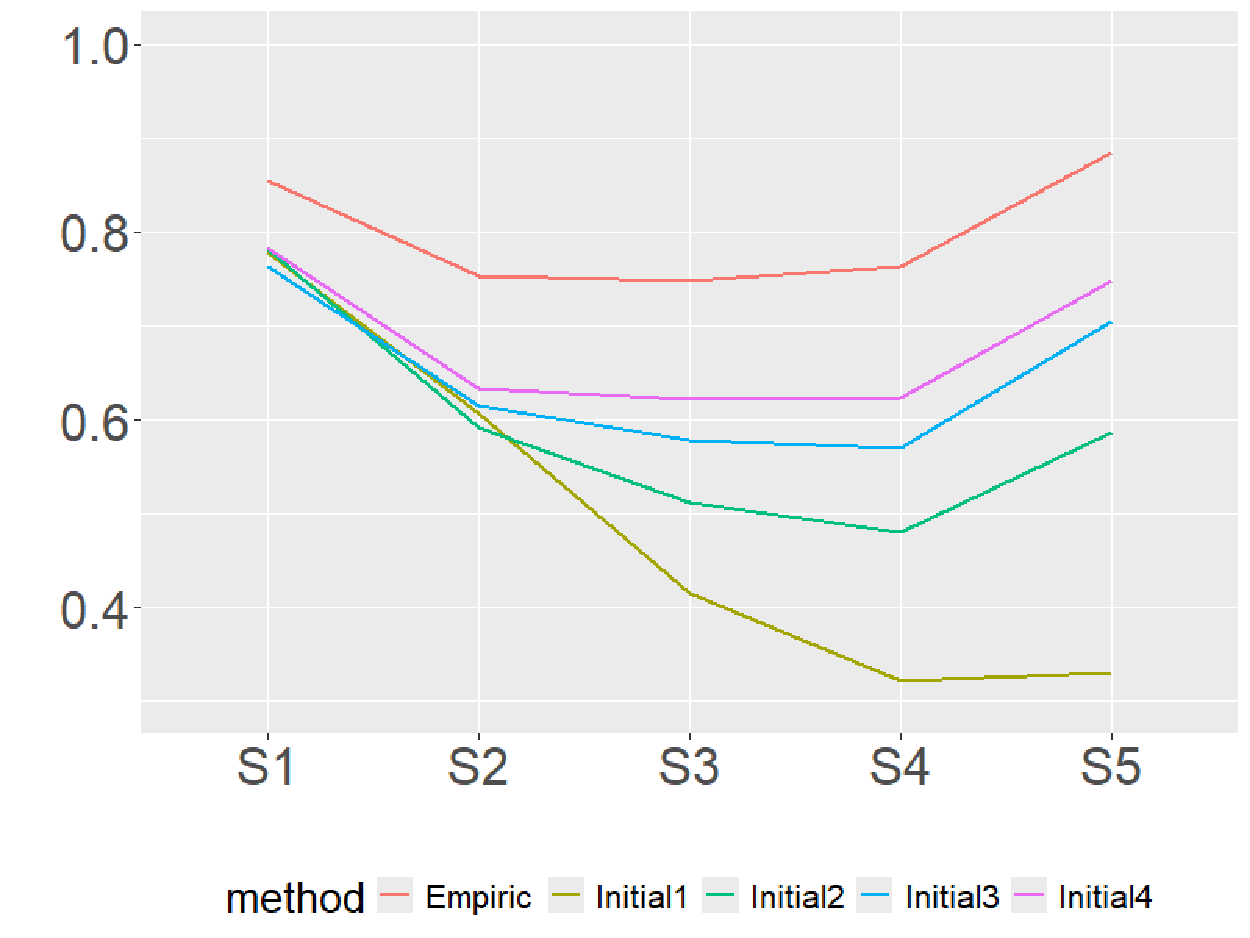}
    \caption{PCS of the Probit method across different scenarios and initial stages, with Empiric results included for reference.}
    \label{fig:initals_probit}
\end{figure}

Simulation results for the PCS of the Empiric method under various initial stages are presented in Appendix~\ref{sup:empiric_initial}. These results indicate that the performance of the Empiric method is largely unaffected by the choice of initial stage. Consequently, we report only the results under Initial1 in Figure~\ref{fig:initals_probit}, illustrating that the performance of this method is less sensitive to initial conditions than those of the two-parameter logistic and probit models.

Figure~\ref{fig:initals_probit} illustrates that the PCS of the Probit method in more challenging scenarios, where the target dose $d^T$ is at higher dose levels, can be substantially low if the initial stage is not sufficiently complex. For instance, in scenarios $S_4$ and $S_5$ under Initial1, fewer than 40\% of trials correctly identify the target dose, a level of performance that is suboptimal for clinical application. This observation is consistent with the findings of Paoletti and Kramar~\parencite{paoletti2009comparison}, who note that \textit{‘The situation of the highest dose being the MTD is not representative of the performance of the methods. The two-parameter model never picks up the right dose when using a likelihood estimator’}. In most investigated scenarios, except for $S_1$, the PCS of the Probit method improves with increased complexity in the initial stage. In scenario $S_1$, where the target dose corresponds to the lowest dose level, the PCS appears relatively insensitive to the complexity of the initial stage, even when the dose-toxicity curve exhibits an undesirable step function.

%However, a larger initial stage, such as Initial4, may allocate an excessive number of patients to Stage 1, reducing the sample size for Stage 2 and ultimately leading to poorer PCS outcomes. This suggests that increased complexity in the initial stage does not always yield optimal results.

The Empiric model consistently outperforms the Probit method across the investigated scenarios, as demonstrated in Paoletti and Kramar~\parencite{altzerinakou2021change}. An often overlooked advantage of the Empiric method is its capacity to mitigate the separation issue in parameter estimation, thereby maintaining stable performance across varying initial stages. However, as demonstrated in Section~\ref{sec:calibration}, this advantage over the Probit and Logistic models may diminish—or even disappear—after dose label calibration.

It is important to note that there is no universal standard for the complexity of the initial stage for the dose-finding studies. The choice should depend on the sample size of the trial while also considering the inherent separation problem. In contrast, for the Empiric method (see Appendix~\ref{sup:empiric_initial}), more complex initial stages provide no added benefit in terms of the proportion of separation or PCS, though they come at the cost of patient outcomes during the initial stage.

The trade-off between the benefit of a smaller proportion of separation and the cost of a larger initial stage for the Probit method is illustrated in Table~\ref{tab:cost_benef}. As shown in Table~\ref{tab:prop_sepa}, in four out of five scenarios, more than 50\% of simulations result in a high degree of separation when the heterogeneity requirement is implemented using Initial1. This issue is substantially mitigated under Initial2. While a more complex initial stage can effectively reduce the separation problem, the associated cost in terms of time and patients should not be overlooked (as shown in Table~\ref{tab:ave_ini}). For instance, with Initial4, the initial stage consumes more than one-quarter of the total sample size on average before the method can proceed in over half of the evaluated scenarios.

%The distribution of required cohorts during the initial stage is provided in Appendix~\ref{sup:cohortsInitialStage}. Given these considerations, we select Initial3 for the subsequent comprehensive simulation results.

\begin{table}[H]
\centering
    \begin{subtable}[h]{0.45\textwidth}
        \centering
        \begin{tabular}{ccccc}
\hline
 & Initial1 & Initial2 & Initial3 & Initial4 \\
\hline
$S_1$ &  0.245  &  0.225 &  0.228  &  0.185 \\
$S_2$ &  0.570  &  0.475 &  0.426  & 0.315 \\
$S_3$ &  0.607  &  0.414  &  0.348  &  0.238 \\
$S_4$ &  0.834  &  0.597  &  0.478  &  0.369 \\
$S_5$ &  0.875  &  0.587  &  0.459  & 0.403 \\
\hline
\end{tabular}
\caption{Proportion of the separation}
\label{tab:prop_sepa}
\end{subtable}
    \hfill
    \begin{subtable}[h]{0.45\textwidth}
        \centering
\begin{tabular}{ccccc}
\hline
 & Initial1 & Initial2 & Initial3 & Initial4 \\
\hline
$S_1$ & 1.40  &   2.42   &  3.12  &   3.95 \\
$S_2$ & 1.86  &   3.01  &   4.05   &  5.06 \\
$S_3$ & 2.29  &   3.72  &   4.89  &   6.00 \\
$S_4$ & 3.00  &   4.58   &  5.87   &  7.01 \\
$S_5$ & 3.52   &  5.44   &  7.00  &   8.44 \\
\hline
\end{tabular}
\caption{Average number of cohorts in the initial stage}
\label{tab:ave_ini}
\end{subtable}
\caption{Benefit and cost of the Probit method under increasingly complex initial stages.}
\label{tab:cost_benef}
\end{table}

\begin{figure}[H]
    \centering
    \includegraphics[scale=0.5]{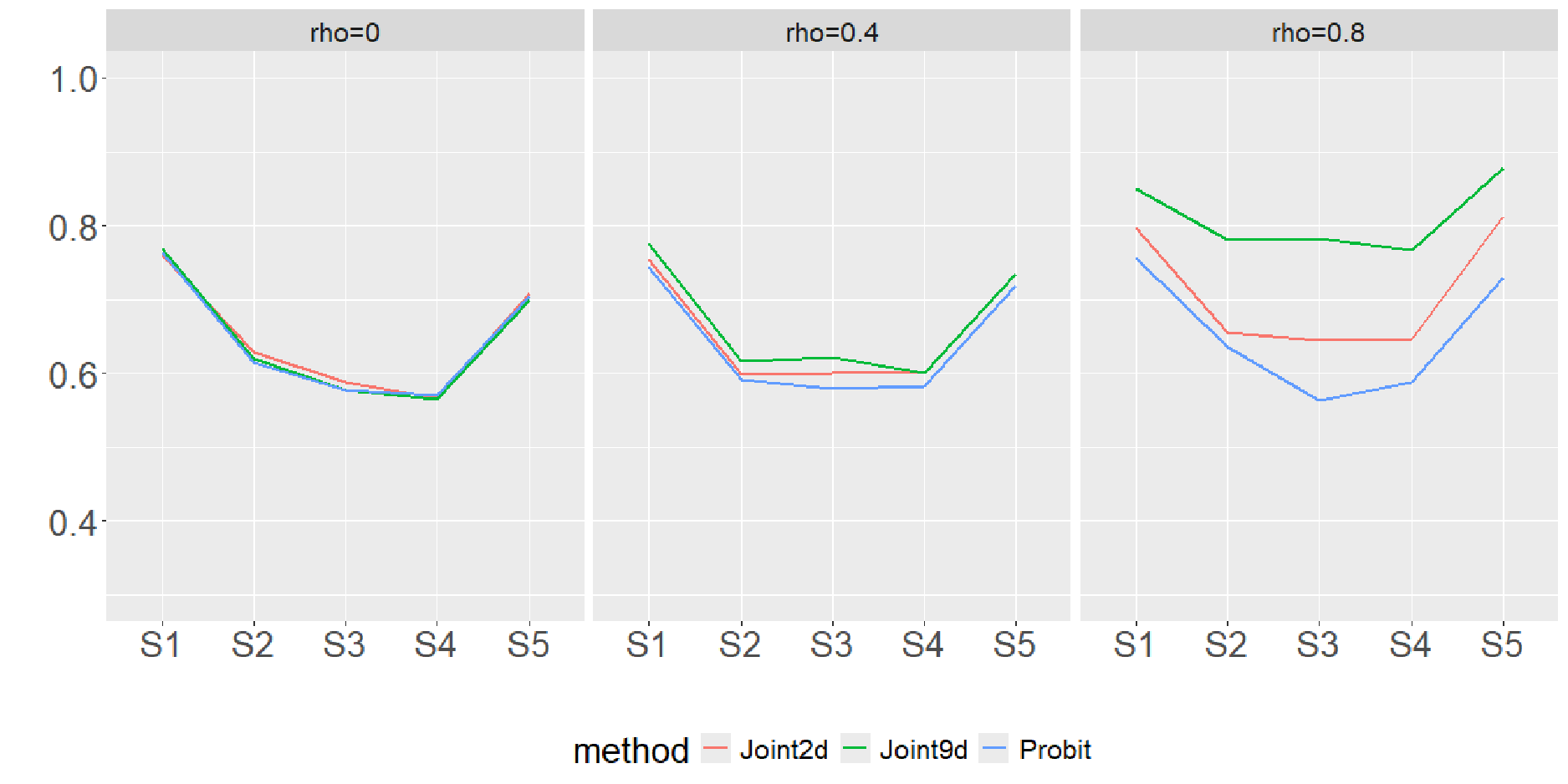}
    \caption{Proportion of correct selections across all scenarios. All three methods were implemented with Initial3 as the initial stage.}
    \label{fig:pcs_all}
\end{figure}

Figure~\ref{fig:pcs_all} presents the PCS results for all three methods across all scenarios under the Initial3 configuration. When biomarker is purely noise ($\rho_b = 0$), the PCS results show negligible differences among these methods, with most cases differing by less than 2\%. However, as biomarker becomes more informative (i.e., $\rho_b = 0.4$ and $\rho_b = 0.8$), significant differences in PCS emerge across the methods. Table~\ref{tab:pcs_sep} presents the PCS across five scenarios ($S_1$–$S_5$) under varying levels of correlation between binary and continuous outcomes ($\rho_b = 0, 0.4, 0.8$). 

Overall, both joint modeling approaches (Joint2d and Joint9d) consistently outperform the baseline Probit method, particularly as the correlation $\rho_b$ increases. Specifically, Joint2d yields a 3–10\% absolute improvement in PCS across scenarios from $\rho_b=0$ to $\rho_b=0.8$, while Joint9d achieves a 10–20\% improvement. This suggests that the joint models are better able to leverage the underlying correlation structure between toxicity and biomarker. In scenario $S_1$, which represents a relatively easy case where all methods perform well, yielding more than 75\% of PCS, the Joint2d and Joint9d methods still achieve incremental improvements of 4–9\%, reaching a PCS of 80\% and 85\%, respectively, at $\rho_b = 0.8$. In contrast, scenario $S_5$ represents the most challenging case, where the benefits of joint modeling are most evident: Joint2d improves PCS from 0.70 to 0.81 (an 11\% gain), and Joint9d from 0.70 to 0.88 (an 18\% gain), as $\rho_b$ increases from 0 to 0.8. In such difficult scenarios, we also focus on the proportion of separations (discussed in Section~\ref{subsec:stuck}).

Simulation results for the proportion of separation, reported alongside PCS values, are summarized in Table~\ref{tab:pcs_sep}. A high separation rate indicates that the model frequently fails to capture the dose-toxicity relationship as the failure of finding the MLE. Under the least informative setting ($\rho_b = 0$), all methods exhibit high separation rates, particularly in the more challenging scenarios $S_4$ and $S_5$, where the separation proportion exceeds 40\%. As the informativeness of biomarker increases (e.g., $\rho_b = 0.8$), both the accuracy of dose selection (PCS) and the stability of the model (separation rate) improve significantly. Among the methods evaluated, Joint9d exhibits the most pronounced reduction in separation. In scenario $S_5$, for instance, the separation rate decreases from 45.6\% to 14.9\%, and falls below 10\% in most scenarios—reaching as low as 7.8\% in $S_4$, 2.8\% in $S_3$, and 3.6\% in $S_2$. These results indicate that incorporating biomarker information is highly effective in mitigating the separation problem across a range of scenarios. In contrast, scenario $S_1$ consistently shows low separation rates across all methods and $\rho_b$ values. This suggests that in simpler settings—where informative data accumulate rapidly at lower dose levels—the anchoring issue has limited influence on the final dose recommendation, as early-stage information at the lowest dose is sufficient to guide correct selection.

\begin{table}[H]
\begin{tabular}{c|ccc|cc|cc}
\hline  & \multicolumn{3}{|c|}{$\rho_b=0$} & \multicolumn{2}{|c|}{$\rho_b=0.4$} & \multicolumn{2}{|c}{$\rho_b=0.8$} \\
\hline  & Probit & Joint2d & Joint9d & Joint2d & Joint9d  & Joint2d & Joint9d \\
\hline S1 & 0.76 (0.225) & 0.76 (0.219) & 0.77 (0.221)  & 0.75 (0.217) & 0.77 (0.231)  & 0.80 (0.193) & 0.85 (0.057) \\
\hline S2 & 0.61 (0.426) & 0.63 (0.402) & 0.62 (0.394) & 0.60 (0.379) & 0.62 (0.342) & 0.66 (0.287)& 0.78 (0.036)\\
\hline S3 & 0.58 (0.330) & 0.59 (0.329) & 0.58 (0.321) & 0.60 (0.286) & 0.62 (0.294)& 0.65 (0.222)& 0.78 (0.028)\\
\hline S4 & 0.57 (0.462) & 0.57 (0.460) & 0.57 (0.442)  & 0.60 (0.402) & 0.60 (0.403)& 0.65 (0.285)& 0.77 (0.074)\\
\hline S5 & 0.7 (0.449) & 0.71 (0.453) & 0.70 (0.456)  & 0.74 (0.441) & 0.74 (0.445) & 0.81 (0.332)& 0.88 (0.149)\\
\hline
\end{tabular}
\caption{Proportion of correct selections across all scenarios, with proportion of separation indicated in brackets. All three methods were implemented with Initial3 as the initial stage.}
\label{tab:pcs_sep}
\end{table}

The issue of equal estimated probabilities, as described in Section~\ref{subsec:eq}, is minimal across all methods and scenarios, with no instance exceeding a 2.5\% occurrence rate (see Table~\ref{tab:pcs_eq}). This phenomenon is largely data-driven and tends to arise under specific conditions—most notably when the target dose $d^T$ is either correctly identified early at a lower dose (e.g., scenario $S_1$) or when there is a general lack of observable treatment response across all dose levels (e.g., scenario $S_5$).

The issue is slightly more prevalent in the Joint9d model under $\rho_b = 0.8$, particularly in scenario $S_1$, where the observed rate of equal estimated probabilities reaches 2.5\%. This increase is likely due to the amplification of patterns in binary outcomes by highly informative biomarker data. When biomarker values provide similar predictive strength as the binary outcomes, they may reinforce existing uncertainties or similarities between dose levels, making it harder to distinguish between competing options.

It is important to emphasize that dose-finding studies prioritize correct dose selection over precise parameter estimation. The presence of equal estimated probabilities does not hinder the trial's progression or its ability to correctly identify the optimal dose. This is ensured by design features such as the no-skipping rule, which enforces logical and structured dose escalation/de-escalation. Consequently, even in the rare cases where estimated probabilities of different doses are similar, all methods remain capable of recommending the correct target dose with high reliability.

\begin{table}[H]
\begin{tabular}{c|ccc|cc|cc}
\hline  & \multicolumn{3}{|c|}{$\rho_b=0$} & \multicolumn{2}{|c|}{$\rho_b=0.4$} & \multicolumn{2}{|c}{$\rho_b=0.8$} \\
\hline  & Probit & Joint2d & Joint9d & Joint2d & Joint9d  & Joint2d & Joint9d \\
\hline S1 &  0.007 & 0.007 & 0.008 & 0.010 & 0.011 & 0.012 & 0.025    \\
\hline S2 & 0      & 0     & 0     & 0     & 0  & 0.003 & 0.007 \\
\hline S3 & 0      &  0    &  0    &  0    &  0  &  0 &  0.007 \\
\hline S4 & 0      & 0     &   0    &  0   &  0  &  0 &  0.006 \\
\hline S5 & 0.002  & 0.004 &  0.003 &  0.005 &   0.003  &   0.003 &   0.007 \\
\hline
\end{tabular}
\caption{Proportion of equal estimated probabilities. All three methods were implemented with Initial3 as the initial stage.}
\label{tab:pcs_eq}
\end{table}

As shown in Table~\ref{tab:prop_earlystopping}, early stopping due to observed toxicities in the initial stage occurs in a small proportion of cases. This is more pronounced in scenario $S_1$, but overall, early termination due to overtoxicity is rare and does not pose a significant concern for the proposed design.

\begin{table}[H]
\begin{tabular}{c c c c}
\hline
$\rho_b$ & 0  & 0.4 & 0.8  \\ 
\hline
$S_1$ & 0.073 & 0.090 & 0.072\\
$S_2$ & 0.004 & 0.012 & 0.005 \\ 
$S_3$ & 0.001 & 0.001 & 0.002 \\ 
$S_4$ & 0 & 0 & 0 \\ 
$S_5$ & 0.001 & 0 & 0 \\ 
\hline
\end{tabular}
\caption{Proportion of early stopping for being overtoxic. All three methods were implemented with Initial3 as the initial stage.}
\label{tab:prop_earlystopping}
\end{table}

\subsection{One-parameter Probit method}\label{sec:one-paraProbit}

\begin{figure}[H]
    \centering
    \includegraphics[scale=0.5]{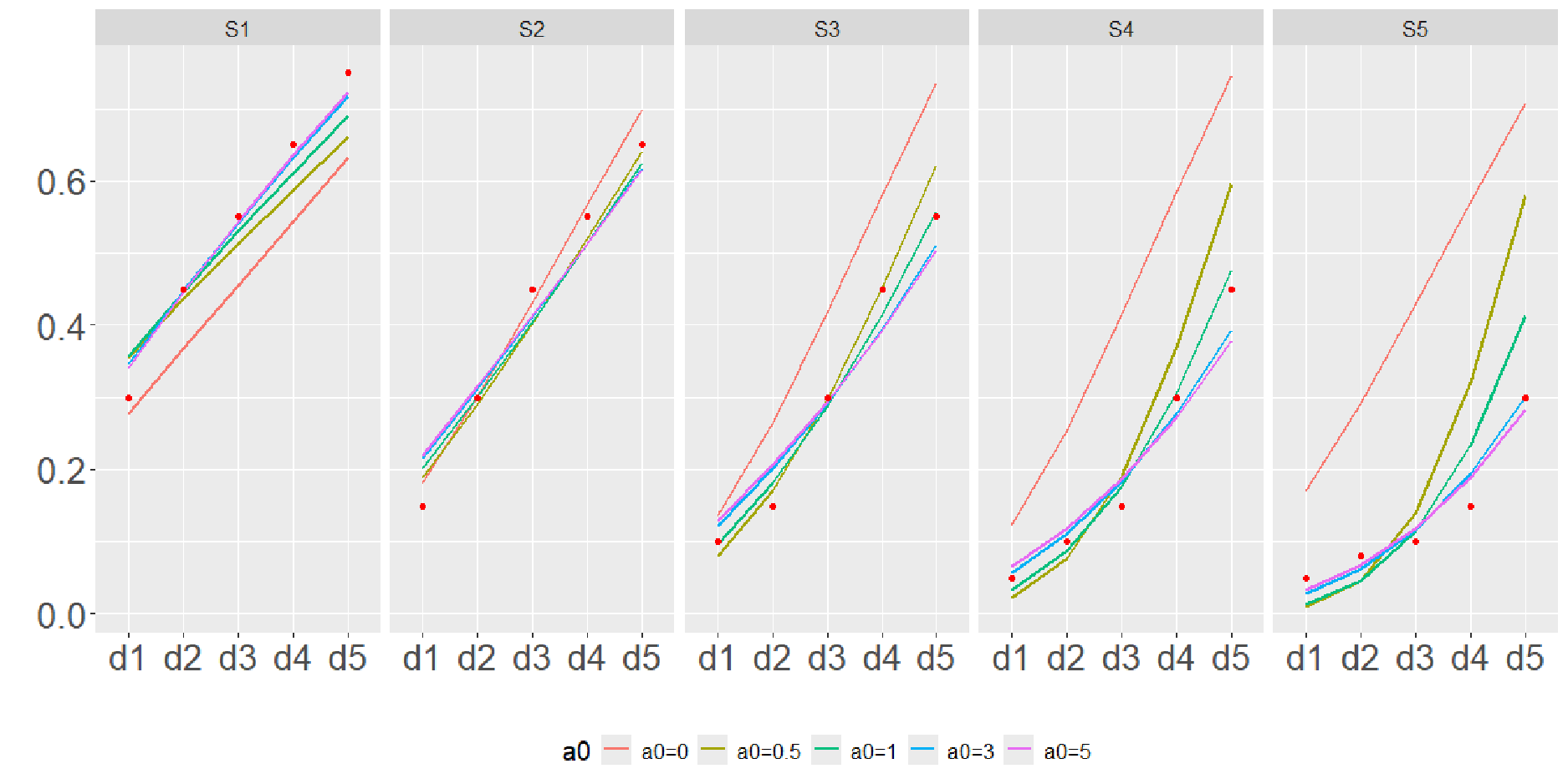}
    \caption{Average estimated probability of toxicity across dose levels under varying values of $a_0$ in the one-parameter Probit model with complete information of doses.}
    \label{fig:ParaEst_1ParaProbit}
\end{figure}

Before using the one-parameter Probit method in the context of dose-finding setting, we first apply the method for all available data for 20 cohorts of patients with all 5 dose levels for parameter estimation. Figure~\ref{fig:ParaEst_1ParaProbit} illustrates the point estimates of $\hat{\pi}({x_j})$ for each dose across different scenarios, obtained using the one-parameter Probit method with varying values of $a_0$. When $a_0=0$, the reference level $x^*$ can dichotomize dose labels between $d_3$ and $d_4$ at $\pi(x^*)=0.5$. This results in doses $d_1$ through $d_3$ being estimated to below 0.5, while $d_4$ and $d_5$ are consistently estimated as more toxic than 0.5 across all scenarios. Consequently, the method performs well only under scenario $S_2$, where the dose-toxicity relationship aligns with the pre-specified skeleton. In contrast, it leads to underestimation in scenario $S_1$ and overestimation in scenarios $S_3$ through $S_5$ as shown in Figure~\ref{fig:One_Para}. 

For all other choices of $a_0$ illustrated in Figure~\ref{fig:One_Para}, the reference dose $x^*$ no longer dichotomizes the dose labels. Instead, it serves as a reasonable upper limit. When $a_0=0.5$, the upper limit $\pi(x^*)=0.691$ leads to an overestimation of the probability of toxicity for higher doses in scenarios where all doses are relatively safe, such as $S_4$ and $S_5$. Furthermore, the reference dose $x^*=0$ causes the estimated probability of toxicity for higher doses, particularly $d_5$, to be biased toward $\pi(x^*)$. Figure~\ref{fig:ParaEst_1ParaProbit} suggests that a small value of $a_0$ is unsuitable. Instead, an appropriately large value of $a_0$ should ensure that $\pi(a_0)\approx 1$ at $x^*=0$, preventing misleading information from locating the curve via the reference dose $x^*$.

%In the context of dose-finding studies (Figure~\ref{fig:ParaEst_1ParaProbit_final}), parameter estimation problem induced by inappropriate choice of $a_0$ is more serious with the sequentially accumulated data. Estimated probability of toxicity under small values of $a_0$ are similar to that without sequential assignment in Figure~\ref{fig:ParaEst_1ParaProbit}. However, the largest value $a_0=5$, which provides appropriate in Figure~\ref{fig:ParaEst_1ParaProbit} fails to work at all, resulting in $\hat\pi(d_j)=1$ for all doses. 

In dose-finding studies, the choice of initial stage has only a marginal impact on the PCS for one-parameter models, including both the empiric (see Section~\ref{sec:initial}) and one-parameter Probit models. This is because issues such as the anchoring problem and the broader separation issues are specific to two-parameter sigmoid models—namely, the two-parameter Probit and Logistic models. We consider the results under Initial1 here.

The dichotomization of dose levels permits accurate identification of the target dose ($d^T$) only when the prespecified skeleton aligns precisely with the true dose–toxicity relationship, as exemplified by scenario $S_3$ with $a_0 = 0$ (see Table~\ref{tab:pcs_probit1para}). In contrast, the model fails to correctly identify $d^T$ under scenarios $S_4$ and $S_5$, as the true target doses ($d_4$ and $d_5$, respectively) are erroneously perceived as overly toxic, with estimated toxicity probabilities exceeding the reference level of $\pi(x^*)=0.5$. Similarly, the specification $a_0 = 0.5$ yields a notably low PCS in $S_5$ due to an overestimation of the toxicity of $d_5$. Conversely, higher values of $a_0$, specifically $a_0 = 1$, $a_0 = 3$, and $a_0 = 5$, demonstrate more robust performance, consistently achieving higher PCS values and accurately identifying the target dose across all examined scenarios. As mentioned in Section~\ref{subsec:reme-1para}, a large value of $a_0$ can easily lead to an extreme argument in the probit model, causing the estimated toxicity probabilities $\pi(d_j)$ to approach 1 for all doses.

Chevret~\parencite{chevret1993continual} investigate the impact of the choice of $a_0$ for the one-parameter logistic model within the Bayesian framework, highlighting that the best choice depends on the situation tested. Moreover, it was noted that for any given situation, the bias of the estimated probability of toxicity was not linear and the recommendation of a fixed value for $a_0$ is somewhat difficult and arbitrary.

%\begin{figure}[H]
%    \centering
%    \includegraphics[scale=0.5]{fig_probit1paraEst_final.eps}
%    \caption{Average value of the estimated probability of toxicity under different choices of $a_0$ in the one-parameter Probit model at the end of dose-finding studies. The implemented initial stage is: Initial2.}
%    \label{fig:ParaEst_1ParaProbit_final}
%\end{figure}

\begin{table}[H]
       \begin{tabular}{c c c c c c}
\hline
$a_0$ & 0  & 0.5 & 1 & 3 & 5 \\ 
\hline
$S_1$ & 0.769 & 0.819 & 0.840 & 0.854 & 0.865 \\
$S_2$ & 0.692 & 0.736 & 0.752 & 0.753 & 0.753 \\ 
$S_3$ & 0.885 & 0.780 & 0.743 & 0.738 & 0.731 \\ 
$S_4$ & 0 & 0.788 & 0.807 & 0.752 & 0.740 \\ 
$S_5$ & 0 & 0.396 & 0.803 & 0.886 & 0.891 \\ 
\hline
\end{tabular}
\caption{PCS of various scenarios under different choices of $a_0$ in the one-parameter probit method. All methods were implemented with Initial1 as the initial stage.}
\label{tab:pcs_probit1para}
\end{table}

%\begin{table}[H]
%\centering
%    \begin{subtable}[h]{0.45\textwidth}
%        \centering
%       \begin{tabular}{c c c c c c}
%\hline
%$a_0$ & 0  & 0.5 & 1 & 3 & 5 \\ 
%\hline
%$S_1$ & 0.769 & 0.819 & 0.840 & 0.854 & 0.865 \\
%$S_2$ & 0.692 & 0.736 & 0.752 & 0.753 & 0.753 \\ 
%$S_3$ & 0.885 & 0.780 & 0.743 & 0.738 & 0.731 \\ 
%$S_4$ & 0 & 0.788 & 0.807 & 0.752 & 0.740 \\ 
%$S_5$ & 0 & 0.396 & 0.803 & 0.886 & 0.891 \\ 
%\hline
%\end{tabular}
%\caption{Initial stage: Initial1.}
%\label{tab:prop_sepa}
%\end{subtable}
%    \hfill
%    \begin{subtable}[h]{0.45\textwidth}
%        \centering
%\begin{tabular}{c c c c c c}
%\hline
%$a_0$ & 0  & 0.5 & 1 & 3 & 5 \\ 
%\hline
%$S_1$ & 0.759 & 0.823 & 0.838 & 0.853 & 0.853 \\
%$S_2$ & 0.703 & 0.735 & 0.747 & 0.750 & 0.749 \\ 
%$S_3$ & 0.868 & 0.785 & 0.765 & 0.724 & 0.717 \\ 
%$S_4$ & 0 & 0.786 & 0.794 & 0.755 & 0.734 \\ 
%$S_5$ & 0 & 0.384 & 0.780 & 0.887 & 0.882 \\ 
%\hline
%\end{tabular}
%\caption{ Initial stage: Initial3.}
%\label{tab:ave_ini}
%\end{subtable}
%\caption{PCS of various scenarios under different choices of $a_0$ in the one-parameter probit method.}
%\label{tab:pcs_probit1para}
%\end{table}

%\begin{figure}[H]
%    \centering
%    \includegraphics[scale=0.6]{fig_probit1Para_pcs.eps}
%    \caption{PCS of various scenarios under different choices of $a_0$ in the one-parameter probit method.}
%    \label{fig:pcs_1ParaProbit}
%\end{figure}

When compared to the results of the two-parameter probit model, as shown in Table~\ref{tab:pcs_sep}, there is a consistent improvement in PCS across various scenarios. Furthermore, no instances of the separation problem are observed as in the empiric model in Appendix~\ref{sup:empiric_initial}.

\begin{figure}[H]
    \centering
    \includegraphics[scale=0.5]{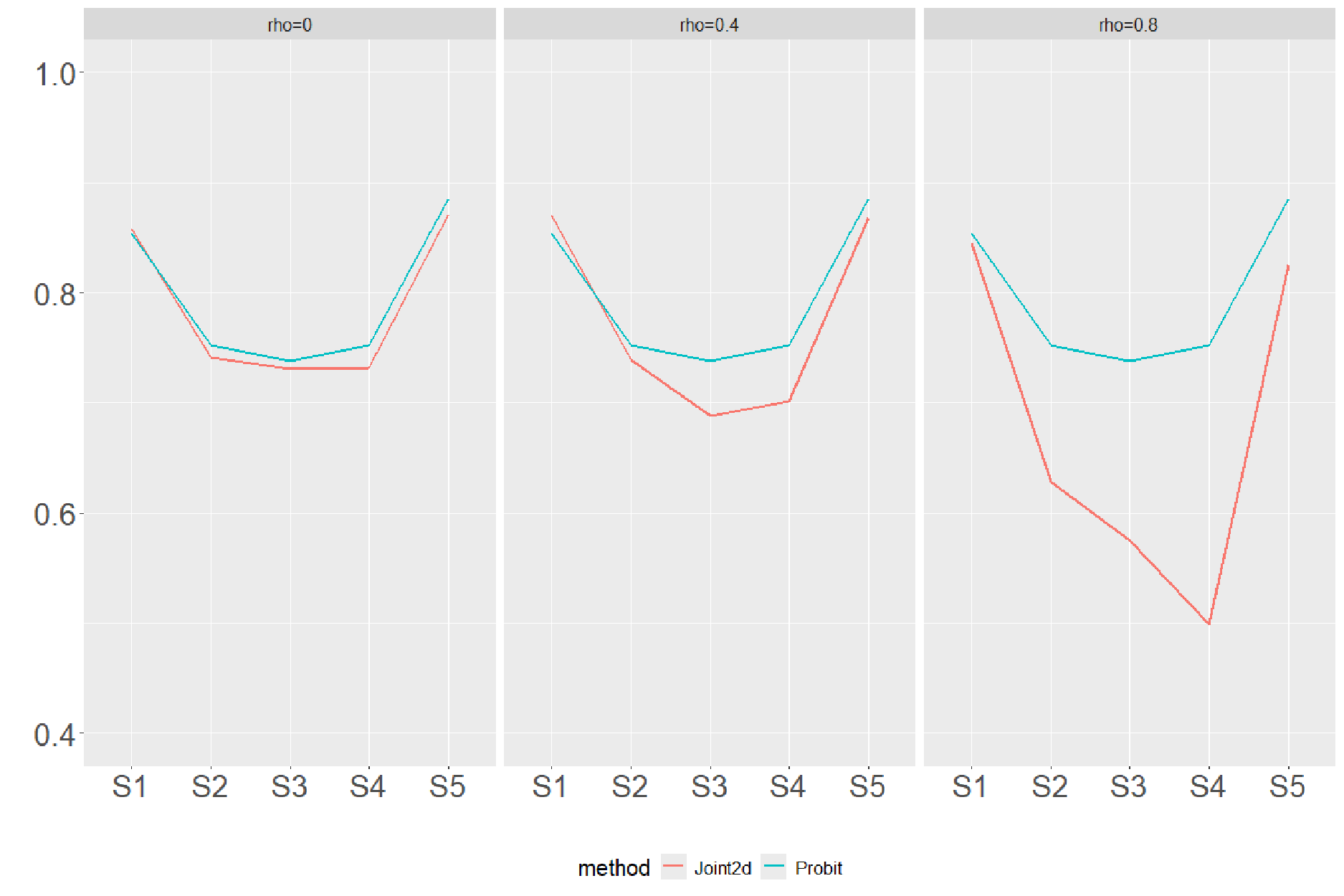}
    \caption{PCS of the Probit and Joint2d method using the one-parameter probit model ($a_0=3$)}
    \label{fig:pcs_1ParaJoint2d}
\end{figure}

Unfortunately, implementing the one-parameter probit model within joint models is not feasible because the marginal effects of interest are derived from the conditional model (see Equation~\ref{eq:cond}). Consequently, the dose label transformation must be based on a scaled intercept rather than a fixed intercept. This scaling accounts for the conditional variance, $1-\rho_b^2$, as shown in Equation~\ref{eq:marginal}. As a result, the intercept used in the joint model is influenced by the value of $\rho_b$. The transformation based on $a_0$ (see Table~\ref{tab:ModelDose}) fails to properly align the dose labels with the skeleton. In other words, joint methods cannot apply the same dose label transformation as the Probit method. Simulation results for the Joint2d method in Figure~\ref{fig:pcs_1ParaJoint2d} illustrate how variations in $\rho_b$ affect PCS outcomes of the Joint2d.

\par

\subsection{Simulation results with calibrated dose labels}\label{sec:calibration}

\begin{figure}[H]
    \centering
    \includegraphics[scale=0.5]{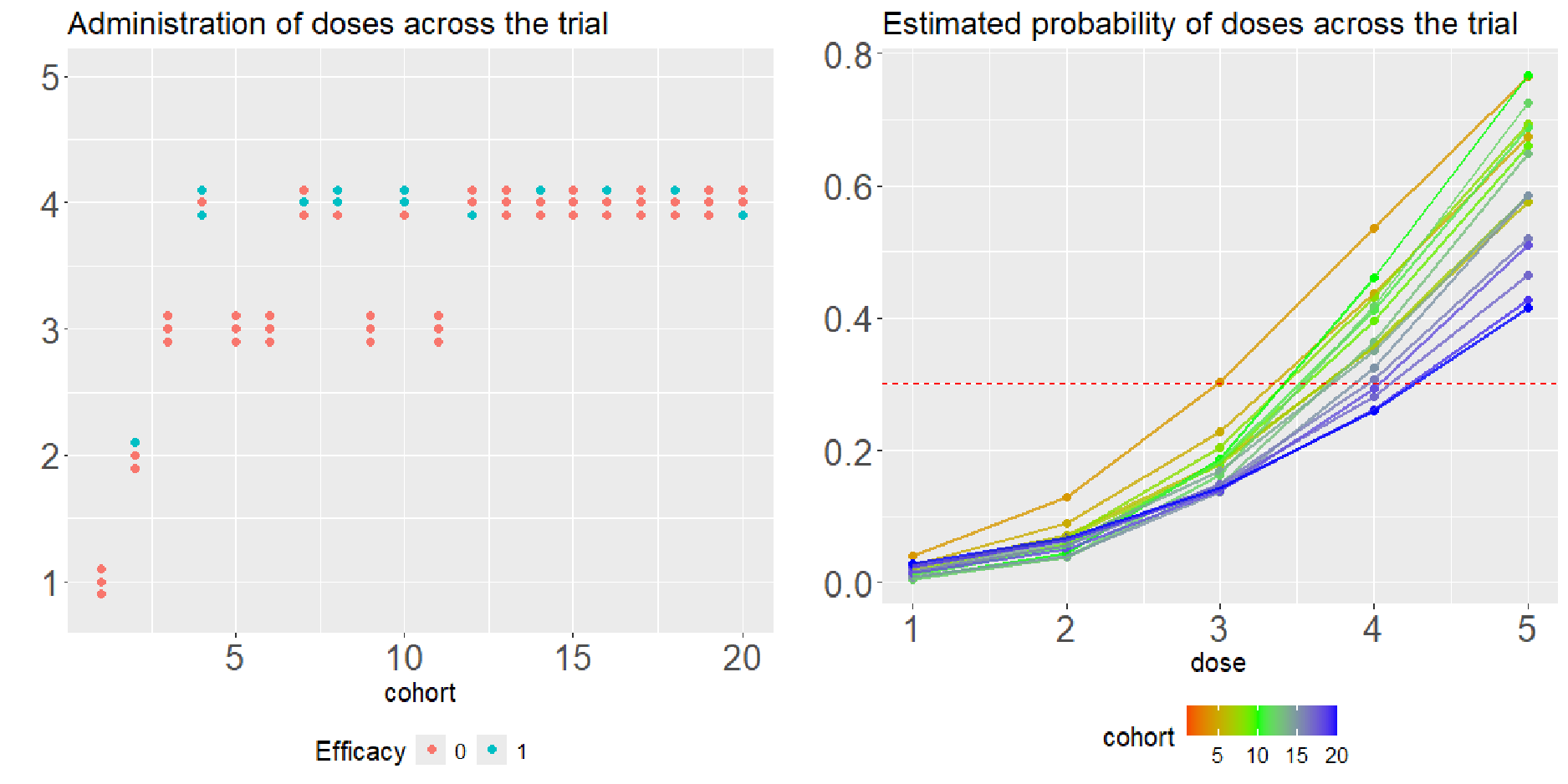}
    \caption{Illustration of the hypothetical trial without the anchoring problem after calibration for dose labels (Initial3).}
    \label{fig:cali_pcs_probit_initial3}
\end{figure}

\begin{figure}[H]
    \centering
    \includegraphics[scale=0.5]{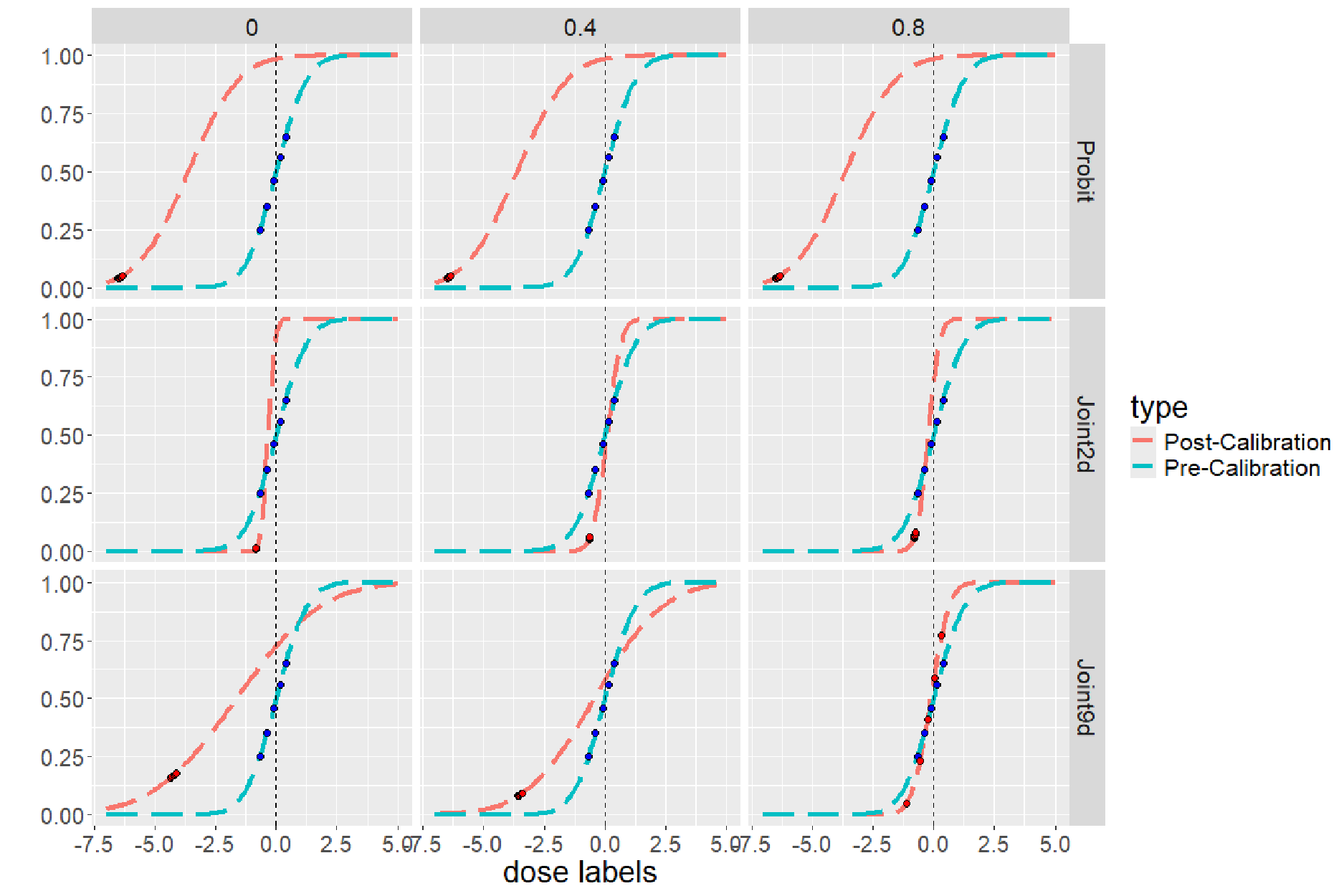}
    \caption{Calibrated model and dose labels.}
    \label{fig:model_after_cali}
\end{figure}

Figure~\ref{fig:cali_pcs_probit_initial3} shows the simulation results for the same example trial presented in Figure~\ref{fig:stuck_initial1_probit}, analyzed with the Probit method before and after dose label calibration. Calibration aligns the dose labels more accurately with the underlying dose-response curve, causing the calibrated labels in this hypothetical trial to cluster toward the lower end of the curve (see Figure~\ref{fig:model_after_cali}). This adjustment resolves the issue of the Probit method becoming anchored at incorrect dose levels, enabling successful escalation to $d_4$. Furthermore, after calibration, the estimated dose-response relationship transitions smoothly as a curve rather than exhibiting a step-function pattern. Thus, calibrating the dose labels can also help mitigate the separation problem observed in Figure~\ref{fig:stuck_initial1_probit}. 

Because Figure~\ref{fig:nonstuck_initial3_all3} demonstrates the influence of a more complex initial stage on estimation results, we additionally applied the same calibrated dose labels under the Initial1 configuration to isolate the effect of calibration itself. The corresponding results, shown in Appendix~\ref{sup:calibration_initial1}, indicate that calibration alone can resolve the separation issue in this individual example.

\begin{figure}[H]
    \centering
    \includegraphics[scale=0.6]{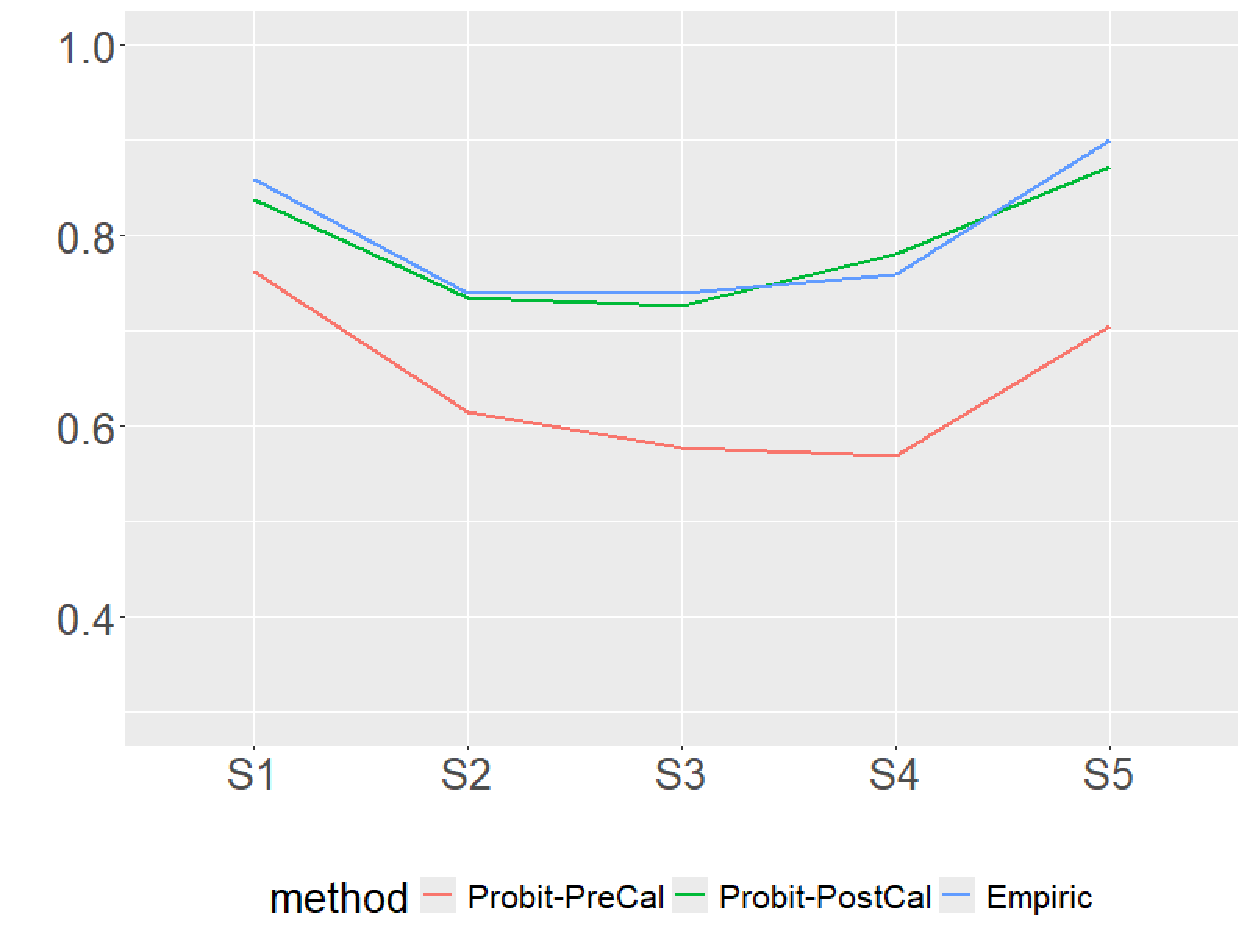}
    \caption{PCS results of the Probit method before calibration (Probit-PreCal) and after calibration (Probit-PostCal). The result of the Empiric method is also included as a reference. All three methods implemented the initial stage: Initial3.}
    \label{fig:calibrate_probit}
\end{figure}

In larger simulations, the PCS results for the Probit method, shown in Figure~\ref{fig:calibrate_probit}, demonstrate substantial improvement after calibration. Following calibration, the performance of the Probit method becomes comparable to that of the Empiric method. Importantly, this finding suggests that the widely accepted conclusion regarding the empiric model’s superior performance—drawn from simulation studies conducted without calibrating the Probit model~\parencite{paoletti2009comparison}—may be incomplete.

Calibration also helps mitigate the separation problem by aligning dose labels more accurately along the sigmoid curve (see the proportion of separation in Table~\ref{tab:pcs_calibration}). This adjustment is particularly impactful in challenging scenarios where the target dose $d^T$ is the highest dose and lower doses exhibit low probabilities of toxicity (as in scenario $S_5$), making early toxic responses less likely. In such cases, calibration distributes dose labels more densely across the lower region of the curve, reflecting lower probabilities $\pi(x_i)$ for these doses. Consequently, the proportion of separation is largely reduced from 44.2\% to less than 1\% (see $S_5$ in Table~\ref{tab:pcs_calibration}). While the Empiric method has often been viewed as superior to the two-parameter Probit model, our findings indicate that, with properly calibrated dose labels, the Probit method can achieve comparable or even superior PCS performance. Although this study does not conduct calibration to the Empiric model, future work exploring dose label optimization in that framework could potentially enhance its performance further.

\begin{table}[H]
\centering
\begin{tabular}{cccc}
\hline
& Empiric & Probit-PreCal & Probit-PostCal   \\
\hline
$S_1$ & 0.86 (0) & 0.76 (0.223)  &  0.84 (0) \\
$S_2$ & 0.74 (0)& 0.62 (0.406)  &  0.73 (0)  \\
$S_3$ & 0.74 (0)& 0.58 (0.322)  &  0.73 (0)  \\
$S_4$ & 0.76 (0)& 0.57 (0.428)  &  0.78 (0.001)  \\
$S_5$ & 0.90 (0)& 0.70 (0.442)  &  0.87 (0.002)  \\
\hline
\end{tabular}
\caption{PCS results of the Probit method before calibration (Probit-PreCal) and after calibration (Probit-
PostCal). The proportion of separation is included in
the bracket.}
\label{tab:pcs_calibration}
\end{table}

Figure~\ref{fig:model_after_cali} presents the model and corresponding dose labels both before and after calibration. Note that the calibration procedure is specific to each method and each value of $\rho_b$.

\begin{figure}[H]
    \centering
    \includegraphics[scale=0.5]{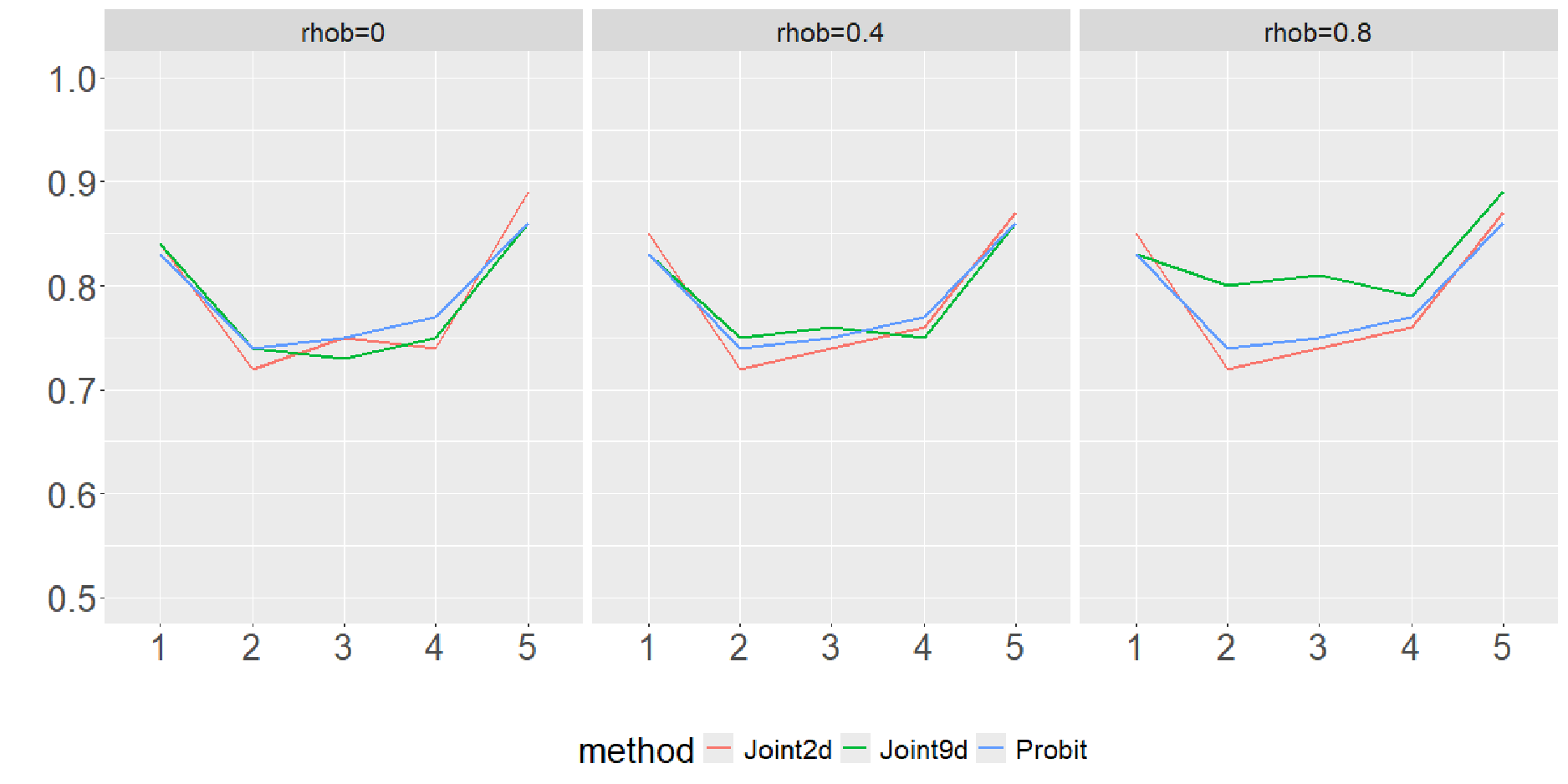}
    \caption{PCS results with calibrated dose labels. All three methods implemented the initial stage: Initial3.}
    \label{fig:pcs_all_cali}
\end{figure}

Figure~\ref{fig:pcs_all_cali} shows the simulation results for all three methods after calibration. The PCS of all methods improves significantly compared to pre-calibration results (see Figure~\ref{fig:pcs_all} and Table~\ref{tab:pcs_sep}). The improvement is most pronounced in the Probit method, where PCS increases by more than 10\% in four out of five scenarios. This highlights the critical role of dose label calibration. For the Joint2d method, the benefit of incorporating biomarker values, as shown in Figure~\ref{fig:pcs_all}, diminishes after calibration. Specifically, the additional information from a single biomarker is insufficient to compensate the added complexity of the model. Similarly, for the Joint9d method, the differences in PCS across $\rho_b$ values diminish after calibration. However, the advantage of incorporating informative biomarker (e.g., $\rho_b=0.8$) remains evident, particularly in scenarios $S_2$–$S_4$. This also reflects the fact that the scenario $S_5$ is a scenario highly sensitive to the separation problem, and this can be largely solved by calibration for dose labels.

Another key advantage of the joint models over the Probit model is their substantial reduction in separation. For the Probit method, calibrating the dose labels significantly mitigates separation, as shown in Table~\ref{tab:pcs_sep_calibration}: in three out of five scenarios, fewer than 1\% of simulations exhibit separation in three out of five scenarios. In the most challenging scenario, $S_5$, separation decreases substantially from the 47\% before calibration to just 8\% after. The joint models show a similarly notable reduction in separation across most scenarios following calibration. Comparing results before and after calibration reveals that poor performance in difficult settings like $S_5$, is largely driven by the anchoring problem—a previously underrecognized issue in the CRM literature. Importantly, many of the separation cases observed prior to calibration lead to correct dose selection once calibration is applied.

\begin{table}[H]
\begin{tabular}{c|ccc|cc|cc}
\hline   & \multicolumn{3}{|c|}{$\rho_b=0$} & \multicolumn{2}{|c|}{$\rho_b=0.4$} & \multicolumn{2}{|c}{$\rho_b=0.8$} \\
\hline  & Probit & Joint2d & Joint9d & Joint2d & Joint9d  & Joint2d & Joint9d \\
\hline S1 & 0.83 (0.001) & 0.84 (0) & 0.84 (0)  & 0.83 (0.000) & 0.84 (0)  & 0.85 (0.002) & 0.83 (0.024) \\
\hline S2 & 0.74 (0.000) & 0.72 (0) & 0.74 (0) & 0.75 (0.000) & 0.69 (0) & 0.72 (0)& 0.80 (0.023)\\
\hline S3 & 0.75 (0.003) & 0.75 (0) & 0.73 (0) & 0.76 (0.004) & 0.74 (0)& 0.74 (0)& 0.81 (0.092)\\
\hline S4 & 0.77 (0.044) & 0.74 (0.002) & 0.75 (0.004)  & 0.75 (0.018) & 0.73 (0.004)& 0.76 (0.038)& 0.79 (0.161)\\
\hline S5 & 0.86 (0.079) & 0.89 (0.004) & 0.86 (0.006)  & 0.86 (0.066) & 0.88 (0.014) & 0.87 (048)& 0.89 (0.210)\\
\hline
\end{tabular}
\caption{PCS results under all scenarios. The proportion of separation is included in the bracket.}
\label{tab:pcs_sep_calibration}
\end{table}

Overall, the anchoring issue in dose-finding studies—or the separation problem in parameter estimation—arises from the limited data typically available early in trials, which is often insufficient to accurately characterize the target dose-response curve. Calibrating the dose labels helps align the investigated dose levels more appropriately along this curve. Our findings indicate that dose label calibration is critical for the performance of likelihood-based approaches in dose-finding studies.

Moreover, the simulation results for the Joint9d method demonstrate that the proposed joint model remains highly robust when informative biomarkers are available (i.e., $\rho_b = 0.8$). In this scenario, dose labels remain sparse even after calibration (see Figure~\ref{fig:model_after_cali}). Nonetheless, the PCS results are quite similar before and after calibration, with some observed improvement. These findings underscore an additional advantage of incorporating biomarkers through joint modeling, as it helps mitigate challenges associated with the problem of separation (see Section~\ref{sec:problem}).

Table~\ref{tab:pcs_eq_calibration} shows the proportion of trials yielding equal estimated probabilities after calibration. Consistent with the pre-calibration results in Table~\ref{tab:pcs_eq}, scenarios $S_1$ and $S_5$ remain more susceptible to this issue than other scenarios. Among the methods analyzed, only the Joint2d and Joint9d approaches exhibit an increased proportion of equal estimates under $\rho_b=0.4$ and $\rho_b=0.8$.

As discussed in Section~\ref{sec:problem}, the occurrence of equal estimated probabilities is better regarded as a phenomenon rather than a fundamental flaw, as it can still reflect meaningful data patterns and support accurate dose recommendations. However, the increase observed under $\rho_b=0.4$ and $\rho_b=0.8$ suggests that high correlation in biomarker information can amplify these patterns, making equal estimated probabilities more frequent. In other words, when the data lack sufficient informativeness, the model may struggle to differentiate between dose levels, when compared to the Probit method.

This highlights an important limitation of dose label calibration: while it can mitigate certain estimation challenges—such as separation—by creating more densely spaced labels, it cannot fully overcome the issues posed by sparse or limited data. In some cases, it can even exacerbate these issues, as the calibrated dose labels become more densely distributed than before calibration (see Figure~\ref{fig:model_after_cali}).

\begin{table}[H]
\begin{tabular}{c|ccc|cc|cc}
\hline  & \multicolumn{3}{|c|}{$\rho_b=0$} & \multicolumn{2}{|c|}{$\rho_b=0.4$} & \multicolumn{2}{|c}{$\rho_b=0.8$} \\
\hline  & Probit & Joint2d & Joint9d & Joint2d & Joint9d  & Joint2d & Joint9d \\
\hline S1 & 0.002  & 0     & 0.002 & 0.042 & 0.002 & 0.035 & 0.049    \\
\hline S2 & 0      & 0     & 0     & 0.006 & 0     & 0.004 & 0.013 \\
\hline S3 & 0.003  & 0     &  0    & 0.010 &  0    & 0.008 &  0.008 \\
\hline S4 & 0      & 0.002 & 0     & 0.008 &  0    & 0.016 &  0.006 \\
\hline S5 & 0.008  & 0.004 & 0.006 & 0.024 & 0.008 & 0.060 &   0.002 \\
\hline
\end{tabular}
\caption{The proportion of equal estimated probability after calibration.}
\label{tab:pcs_eq_calibration}
\end{table}

\section{Discussion}\label{sec:discussion}

In applying a likelihood-based estimation method, it is essential to prevent parameter estimates from reaching the boundaries of the parameter space and ensure that the likelihood function is not monotonic. Although the well-established heterogeneity requirement, ensuring variability in patient responses, is designed to address these challenges, our findings show that simply meeting this condition is insufficient to guarantee the existence of MLEs. In particular, when lower doses exhibit no toxicity, estimation issues such as the separation problem may still arise, leading to a step function in the estimated probability and potentially anchoring the trial at lower doses. This highlights the need for additional strategies to ensure robust parameter estimation. Furthermore, depending on the data, equal estimated probabilities may occur, adding further complexity to the estimation process.

Among the extensive literature on CRM, various simulation studies explore the selection of key design parameters for both Bayesian and likelihood approaches. Our work highlights the inherent limitations in early dose-finding studies from the perspective of data configuration. While our findings align with prior research, we also provide deeper insights into the challenges encountered in more complex scenarios, emphasizing the importance of careful application of model-based designs in dose-finding studies.

One potential approach to mitigating these estimation challenges involves refining the initial stage of model fittingby requiring more occurrences before initiating the model fitting process. In the case of one-parameter models, the choice of the fixed intercept $a_0$ in the one-parameter Probit model plays a crucial role in performance. Our findings align with those of Chevret~\parencite{chevret1993continual}. However, the one-parameter model with a fixed intercept cannot be directly extended to the joint method due to the lack of dose label transformations compatible with repeated biomarker measurements. Dose label calibration, which aligns dose labels more accurately along the dose-response curve, was shown to effectively enhance PCS and reduce estimation issues in both the Probit and joint methods. Despite these improvements, certain scenarios still faced challenges due to the inherent limitations of the dataset, highlighting the constraints of likelihood-based approaches in early-phase dose-finding studies.

Ideally, when data are sufficiently informative, parameter estimates remain consistent regardless of the optimization algorithm used. However, with limited data, different algorithms can produce substantially divergent estimates, potentially leading to inconsistent dosing decisions.

In the absence of a standardized optimization approach, we utilized the R package \textit{optimx}\parencite{optimx2014}, a flexible optimization wrapper offering access to multiple methods available in R. This allowed us to systematically test various algorithms during the MLE procedure.

We also investigated optimization strategies—such as algorithm selection and starting value initialization (results not shown)—to address numerical instability during estimation. While algorithm performance varied across individual trials, no single algorithm fully resolved the fundamental challenges posed by sparse data. Likewise, incorporating a starting value selection procedure did not alleviate the estimation difficulties. These findings highlight a critical limitation of the frequentist framework, which is heavily dependent on appropriate dose labels and remains especially sensitive to data configurations.

The proposed joint method proves to be more robust than traditional approaches, showing consistent performance regardless of the initial stage and dose label calibration. It achieves improved PCS results and fewer estimation issues overall, making it a promising tool in dose-finding studies. However, to fully support the implementation of the joint method framework, further comprehensive evaluation is necessary to address remaining estimation challenges, particularly in scenarios with sparse data. While the current frequentist approach offers valuable insights, adopting a Bayesian framework could be a more effective solution. Bayesian methods, with their ability to incorporate prior information and handle uncertainty more effectively, may provide additional stability and flexibility, ultimately enhancing the robustness of the joint method in real-world clinical applications. Thus, future work should focus on integrating Bayesian techniques for both Probit method and joint methods, ensuring that it can consistently deliver accurate and reliable dose estimates across a broader range of scenarios.

\section*{Acknowledgement}
This project has received funding from the European Union’s Horizon 2020 research and innovation programme under grant agreement No. 965397. This report is independent research supported by the National Institute for Health and Care Research (NIHR Advanced Fellowship, Dr
Pavel Mozgunov, NIHR300576)). PM and TJ also received funding from the UK Medical Research Council (\texttt{MC\_UU\_00002/14}, \texttt{MC\_UU\_00002/19}, and \texttt{MC\_UU\_00040/03})). Infrastructure support is acknowledged from Cancer Research UK, the NIHR Cambridge Biomedical Research Centre (BRC-1215-20014) and Cambridge Experimental Cancer Medicine Centre. For the purpose of open access, the author has applied a Creative Commons Attribution (CC BY) license to any Author Accepted Manuscript version arising.

\clearpage
\begin{appendices}
\section{Flowchart of the two-stage design}\label{sup:flowchart}

\begin{figure}[H]
    \centering
    \caption{Flowchart of the two-stage design}
    \label{sup:flowchart}
\end{figure}

\clearpage

\section{Derivation for the Joint9d method}
\label{sup:joint9d}

%The Joint9d method models the joint probability of binary $Y_b$ and all biomarker values ($Y_{c,1},\ Y_{c,2},\ ...,\ Y_{c,8}$). Here, the assumption about autocorrelation for adjacent biomarker values is implemented via $\mathbf{\Sigma_c}$. The parameter $\rho_b$ to indicate the correlation between $Y^*_b$ and $Y_{c,8}$, the biomarker measurement most closely associated with $Y_b$.

The Joint9d method models the joint probability for individual $i$ at administered dose $x_j$ of the binary variable $Y_{bi}$ and all $Y_{ci,t}$, $t\in (1,\ ...,\ 8)$. For the correlation $\rho_b$ between $Y_b$ and $Y_{c,8}$, we assume $\rho_b\ge0$. To account for negative correlations between outcomes in higher dimensions, the sign of the correlation is introduced explicitly, avoiding ambiguity in interpretation.

Following the framework of factorization above, we are now interested in the marginal model for $\mathbf{Y}_{ci}$, which is multivariate normally distributed, and a conditional model for $Y_{bi}$ given $\mathbf{Y}_{ci}$. The derived conditional distribution is:

\begin{equation}
E(Y_{bi}|\mathbf{Y_{ci}}=\mathbf{y_{ci}})=\Phi\left({\mu_{bi} + \mathbf{\tau}(\mathbf{Y_{ci}} - \mathbf{\mu_{ci}})}\right)\text{ and } \mathbf{Y_{ci}}=\mathbf{\mu_{ci}} +\mathbf{\epsilon_{ci}}.
\label{eq:Joint9d_cond}
\end{equation}

where $\mu_{bi}=\beta_{b0}+\text{exp}(\beta_{b1})\cdot x_j$, ${\mu_{ci,t}}=\beta_{c0}+\beta_{c1}\cdot x_j+a_t\cdot {t}$ ($t\in {1,\ ...,\ 8}$), and $\mathbf{\epsilon_{ci}}\sim \mathbf{N}_8 \left(\mathbf{0},\ \mathbf{\Sigma_c}\right)$. For biomarker values, $\beta_{c0}$ is the intercept, $\beta_{c1}$ indicates the dose-response relationship, and $a_t$ captures the change of mean values $\mu_t$ with time. Corresponding parameters are expressed as $\boldsymbol{\theta}=(\beta_{b0},\ \beta_{b1},\ \beta_{c0},\ \beta_{c1},\ a_t,\ \tau,\ \sigma_c)$.

\begin{equation}
\begin{pmatrix}
Y_{ci,1} \\
Y_{ci,2} \\
Y_{ci,3} \\
Y_{ci,4} \\
Y_{ci,5} \\
Y_{ci,6} \\
Y_{ci,7} \\
Y_{ci,8} \\
Y^*_{bi}
\end{pmatrix}
\sim \mathcal{N} \left( 
\begin{pmatrix}
\mu_{ci,1} \\
\mu_{ci,2} \\
\mu_{ci,3} \\
\mu_{ci,4} \\
\mu_{ci,5} \\
\mu_{ci,6} \\
\mu_{ci,7} \\
\mu_{ci,8} \\
\mu^*_{bi}
\end{pmatrix}, 
\begin{pmatrix}
 \sigma_c^2 & \rho_c\sigma_c^2 & \rho_c^2\sigma_c^2 &  \rho_c^3\sigma_c^2  &  \rho_c^4\sigma_c^2 &  \rho_c^5\sigma_c^2  &  \rho_c^6\sigma_c^2  &  \rho_c^7\sigma_c^2 & -{\rho_b^8}\sigma_c  \\
  \rho_c\sigma_c^2  & \sigma_c^2 & \rho_c\sigma_c^2 & \rho_c^2\sigma_c^2 &  \rho_c^3\sigma_c^2  &  \rho_c^4\sigma_c^2 &  \rho_c^5\sigma_c^2  &  \rho_c^6\sigma_c^2  &  -\rho_b^7\sigma_c  \\
  \rho_c^2\sigma_c^2 & \rho_c\sigma_c^2 & \sigma_c^2 & \rho_c\sigma_c^2  & \rho_c^2\sigma_c^2 &  \rho_c^3\sigma_c^2  &  \rho_c^4\sigma_c^2 &  \rho_c^5\sigma_c^2  &  -\rho_b^6\sigma_c  \\ 
   \rho_c^3\sigma_c^2 &  \rho_c^2\sigma_c^2 & \rho_c\sigma_c^2  & \sigma_c^2 & \rho_c\sigma_c^2  & \rho_c^2\sigma_c^2 &  \rho_c^3\sigma_c^2  &  \rho_c^4\sigma_c^2 &  -\rho_b^5\sigma_c   \\
   \rho_c^4\sigma_c^2 &  \rho_c^3\sigma_c^2 &  \rho_c^2\sigma_c^2 & \rho_c\sigma_c^2  & \sigma_c^2 & \rho_c\sigma_c^2  & \rho_c^2\sigma_c^2 &  \rho_c^3\sigma_c^2  &  -\rho_b^4\sigma_c \\
   \rho_c^5\sigma_c^2 &   \rho_c^4\sigma_c^2 &  \rho_c^3\sigma_c^2 &  \rho_c^2\sigma_c^2 & \rho_c\sigma_c^2  & \sigma_c^2 & \rho_c\sigma_c^2  & \rho_c^2\sigma_c^2 &  -\rho_b^3\sigma_c \\
   \rho_c^6\sigma_c^2 & \rho_c^5\sigma_c^2 & \rho_c^4\sigma_c^2 &  \rho_c^3\sigma_c^2 &  \rho_c^2\sigma_c^2 & \rho_c\sigma_c^2  & \sigma_c^2 & \rho_c\sigma_c^2  & -\rho_b^2\sigma_c \\
  \rho_c^7\sigma_c^2  & \rho_c^6\sigma_c^2   & \rho_c^5\sigma_c^2 &   \rho_c^4\sigma_c^2 &  \rho_c^3\sigma_c^2 &  \rho_c^2\sigma_c^2 & \rho_c\sigma_c^2  & \sigma_c^2 & -\rho_b\sigma_c   \\
   -{\rho_b^8}\sigma_c & -\rho_b^7\sigma_c  & -\rho_b^6\sigma_c & -\rho_b^5\sigma_c & -\rho_b^4\sigma_c &  -\rho_b^3\sigma_c &  -\rho_b^2\sigma_c & -\rho_b\sigma_c  & 1  \\
\end{pmatrix}
\right)
\end{equation}

The conditional distribution of $Y_{bi}$ given $Y_{ci}$ is

%\begin{equation}

\begin{align}
E(Y_{bi}|\mathbf{Y_{ci}}=  
( \mu_{bi} + 
\left[\begin{array}{c}
-\frac{(-\rho_b^8+\rho_b^7\rho_c)}{\sigma_c(\rho_c^2-1)}  \\ 
-\frac{(-\rho_b^7-\rho_b^7\rho^2+\rho_b^8\rho+\rho_b^6\rho_c)}{\sigma_c(\rho_c^2-1)}  \\ 
-\frac{(-\rho_b^6-\rho_b^6\rho_c^2+\rho_b^7\rho_c+\rho_b^5\rho_c)}{\sigma_c(\rho_c^2-1)} \\
-\frac{(-\rho_b^5-\rho_b^5\rho_c^2+\rho_b^6\rho_c+\rho_b^4\rho_c-\rho_b^5-\rho_b^5\rho_c^2+\rho_b^6\rho_c+\rho_b^4\rho_c)}{\sigma_c(\rho_c^2-1)} \\ 
-\frac{(-\rho_b^4-\rho_b^4\rho_c^2+\rho_b^5\rho_c+\rho_b^3\rho_c)}{\sigma_c(\rho_c^2-1)} \\ 
-\frac{(-\rho_b^3-\rho_b^3\rho_c^2+\rho_b^4\rho_c+\rho_b^2\rho_c)}{\sigma_c(\rho_c^2-1)} \\
-\frac{(-\rho_b^2-\rho_b^2\rho_c^2+\rho_b^3\rho_c+\rho_b\rho_c )}{\sigma_c(\rho_c^2-1)} \\
-\frac{(-\rho_b+\rho_b^2\rho_c)}{\sigma_c(\rho_c^2-1)}
\end{array}\right]^{T}
\left[\begin{array}{c}
(y_{ci,1} - \mu_{ci,1}) \\
(y_{ci,2} - \mu_{ci,2}) \\
(y_{ci,3} - \mu_{ci,3}) \\
(y_{ci,4} - \mu_{ci,4}) \\
(y_{ci,5} - \mu_{ci,5}) \\
(y_{ci,6} - \mu_{ci,6}) \\
(y_{ci,7} - \mu_{ci,7}) \\
(y_{ci,8} - \mu_{ci,8})
\end{array}\right],\ \Sigma_{b*|c}),
\label{eq:joint9d_cond}
\end{align}
%\end{equation}    

where $\mu_{bi}=\beta_{b0}+\text{exp}(\beta_{b1})\cdot x_j$ and 
\begin{multline}
\Sigma_{b*|c}=
     (1-(-\rho_b^{16} -\rho_b^{14} -\rho_b^{12} -\rho_b^{10} -\rho_b^8 -\rho_b^6-\rho_b^4 -\rho_b^2 - \rho_b^{14}\rho_c^2 - \rho_b^{12}\rho_c^2 \\  - \rho_b^{10}\rho_c^2 - \rho_b^{8}\rho_c^2 - \rho_b^{6}\rho_c^2 - \rho_b^{4}\rho_c^2 + 2\rho_b^{15}\rho_c + 2\rho_b^{13}\rho_c + 2\rho_b^{11}\rho_c +
     \\ 2\rho_b^{9}\rho_c + 2\rho_b^{7}\rho_c + 2\rho_b^{5}\rho_c + 2\rho_b^{3}\rho_c)
     \label{eq:joint9d_condvar}
\end{multline}

The reparameterization is derived from the marginal effects outlined in Equation~\ref{eq:joint9d_cond} and the conditional variance specified in Equation~\ref{eq:joint9d_condvar}. That is $\beta^*_{b0}=\beta_{b0}\cdot\sqrt{\Sigma_{b*|c}}$ and $\beta^*_{b1}=\text{exp}(\beta_{b1})\cdot\sqrt{\Sigma_{b*|c}}$.

\clearpage

\section{Constraints for choices of parameters}\label{sup:cond_var}

Results for $\sigma_c = 1$ are presented, as its value influences the conditional variance and the positive definiteness of the covariance matrix.

\begin{figure}[H]
    \centering
    \caption{Conditional variance under different choices of parameter $\rho_b$ and $\rho_c$ ($\sigma_c=1$).}
    \label{fig:condtional_variance}
\end{figure}

\begin{figure}[H]
    \centering
    \caption{Positive definiteness of a variance-covariance matrix under different $\rho_b$ and $\rho_c$, 0 and 1 indicate positive definiteness and non-positive definiteness ($\sigma_c=1$).}
    \label{fig:positive_definite}
\end{figure}

\clearpage

\section{No anchoring problem with Initial2}\label{sup:Initial2}

\begin{figure}[H]
    \centering
    \caption{Illustration of the hypothetical trial without the anchoring problem after implementing an alternative initial stage: Initial2. Stage 1 includes four cohorts and Stage 2 applies the Probit method based on the two-parameter probit model.}
    \label{fig:nostuck_initial2}
\end{figure}

\clearpage

\section{PCS of the Empiric method under different initial stages}
\label{sup:empiric_initial}

\begin{figure}[H]
    \centering
    \caption{PCS of the Empiric method under different scenarios with different initial stages.}
    \label{fig:initals_empiric}
\end{figure}

\begin{table}[H]
\centering
    \begin{subtable}[h]{0.45\textwidth}
        \centering
        \begin{tabular}{ccccc}
\hline
 & Initial1 & Initial2 & Initial3 & Initial4 \\
\hline
$S_1$ &  0  &  0 &  0  &  0 \\
$S_2$ &  0  &  0 &  0  &  0 \\
$S_3$ &  0  &  0 &  0  &  0 \\
$S_4$ &  0  &  0 &  0  &  0 \\
$S_5$ &  0  &  0 &  0  &  0 \\
\hline
\end{tabular}
\caption{Proportion of the separation}
\label{tab:prop_sepa_empiric}
\end{subtable}
    \hfill
    \begin{subtable}[h]{0.45\textwidth}
        \centering
\begin{tabular}{ccccc}
\hline
 & Initial1 & Initial2 & Initial3 & Initial4 \\
\hline
$S_1$ & 1.40  &   2.42   &  3.12  &   3.95 \\
$S_2$ & 1.86  &   3.01  &   4.05   &  5.06 \\
$S_3$ & 2.29  &   3.72  &   4.89  &   6.00 \\
$S_4$ & 3.00  &   4.58   &  5.87   &  7.01 \\
$S_5$ & 3.52   &  5.44   &  7.00  &   8.44 \\
\hline
\end{tabular}
\caption{Average cohorts required in the initial stage}
\label{tab:ave_ini_empiric}
\end{subtable}
\caption{The benefit and cost for the Empiric with more complicated initial stage}
\label{tab:cost_benef_empiric}
\end{table}

\clearpage

%\section{Distribution of the number of cohorts required under different initial stages}\label{sup:cohortsInitialStage}

%\begin{figure}[H]
%    \centering
%    \includegraphics[scale=0.5]{fig_distInitialCohort.eps}
%    \caption{Distribution of the required cohorts in the Probit method with different initial stages.}
%    \label{fig:dist_initals}
%\end{figure}

%\clearpage

\section{Simulation result by the Probit method with calibrated lables with Initial1}\label{sup:calibration_initial1}

\begin{figure}[H]
    \centering
    \caption{Illustration of the hypothetical trial without the anchoring problem after calibration for dose labels (Initial1).}
    \label{fig:cali_individual}
\end{figure}

\begin{figure}[H]
    \centering
    \caption{PCS results of the Probit method before calibration (Probit-PreCal) and after calibration (Probit-PostCal). The result of the Empiric method is also included as a reference. All three methods implemented the initial stage: Initial1.}
    \label{fig:cali_pcs_probit_initial1}
\end{figure}

\begin{table}[H]
\centering
\begin{tabular}{cccc}
\hline
& Empiric & Probit-PreCal & Probit-PostCal   \\
\hline
$S_1$ & 0.86 (0) & 0.78 (0.229)  &  0.85 (0) \\
$S_2$ & 0.74 (0) & 0.61 (0.543)  &  0.75 (0)  \\
$S_3$ & 0.74 (0) & 0.42 (0.589)  &  0.73 (0)  \\
$S_4$ & 0.76 (0) & 0.32 (0.803)  &  0.79 (0)  \\
$S_5$ & 0.90 (0) & 0.33 (0.870)  &  0.86 (0.002)  \\
\hline
\end{tabular}
\caption{PCS results of the Probit method before calibration (Probit-PreCal) and after calibration (Probit-
PostCal). The proportion of separation is included in
the bracket.}
\end{table}

\clearpage

\section{Calibrated dose labels}\label{sup:calibration}

\begin{table}[H]
\begin{tabular}{cc|ccc|cc|cc}
\hline  & &\multicolumn{3}{|c|}{$\rho_b=0$} & \multicolumn{2}{|c|}{$\rho_b=0.4$} & \multicolumn{2}{|c}{$\rho_b=0.8$} \\
\hline  & Pre-Calibration & Probit & Joint2d & Joint9d & Joint2d & Joint9d  & Joint2d & Joint9d \\
\hline $d_1$ &  -0.675 & -6.514 & -0.854 & -4.334 & -0.649 & -3.572 & -0.809  & -1.119 \\
\hline $d_2$ & -0.385  & -6.469 & -0.846 & -4.281   & -0.640  & -3.525 & -0.791 & -0.569 \\
\hline $d_3$ & -0.105 & -6.423 &  -0.838 & -4.227  & -0.630 & -3.477 & -0.773  &  -0.259 \\
\hline $d_4$ & 0.151 & -6.378 & -0.829 & -4.173  & -0.620   & -3.429 &  -0.755 &  0.017 \\
\hline $d_5$ & 0.385 & -6.331 & -0.821  & -4.118  & -0.610 & -3.381 & -0.736  &  0.327  \\
\hline
\end{tabular}
\caption{Dose labels before and after calibration, with calibration applied using a method-specific and $\rho_b$-specific approach.}
\label{tab:labels_calibration}
\end{table}

\section{Estimation problem with and without algorithm selection}\label{sup:algorithm}

\begin{table}[H]
\begin{tabular}{c|cc|cc}
\hline  & \multicolumn{2}{|c|}{Separation} & \multicolumn{2}{|c}{Equal estimated probability}  \\
\hline  & nlminb & best  & nlminb & best \\
\hline S1 &  0.001 & 0 & 0.059 & 0.065    \\
\hline S2 & 0  & 0.001 & 0   & 0.001  \\
\hline S3 & 0.003 &  0.002 &  0.003    &  0  \\
\hline S4 & 0.044 & 0.061 &   0  &  0  \\
\hline S5 & 0.079  & 0.163 &  0.008   &  0.008 \\
\hline
\end{tabular}
\caption{The proportion of separation and the proportion of equal estimated probability for the Probit method with and without the algorithm selection. The column `best' represents results with the algorithm selection selection.}
\label{tab:pcs_sep_algorithm_probit}
\end{table}

\begin{table}[H]
\begin{tabular}{c|cc|cc|cc}
\hline  & \multicolumn{2}{|c|}{$\rho_b=0$} & \multicolumn{2}{|c|}{$\rho_b=0.4$} & \multicolumn{2}{|c}{$\rho_b=0.8$} \\
\hline  & nlminb & best & nlminb & best &  nlminb & best \\
\hline S1 &  0.001 & 0 & 0 & 0.002 & 0.003 & 0.003    \\
\hline S2 & 0.001  & 0 & 0   & 0 & 0.003   & 0  \\
\hline S3 & 0.009 &  0.001 &  0    &  0.008 &  0.008   &  0.006  \\
\hline S4 & 0.095 & 0.036 &   0.003  &  0.088 &  0.070  &  0.051  \\
\hline S5 & 0.224  & 0.082 &  0.005   &  0.161 &   0.133  &   0.061 \\
\hline
\end{tabular}
\caption{The proportion of separation for the Joint2d method with and without the algorithm selection. The column `best' represents results with the algorithm selection selection.}
\label{tab:pcs_sep_algorithm_joint}
\end{table}

\begin{table}[H]
\begin{tabular}{c|cc|cc|cc}
\hline  & \multicolumn{2}{|c|}{$\rho_b=0$} & \multicolumn{2}{|c|}{$\rho_b=0.4$} & \multicolumn{2}{|c}{$\rho_b=0.8$} \\
\hline  & nlminb & best &  nlminb & best &  nlminb & best \\
\hline S1  & 0.063 & 0.089  & 0.072 & 0.123  & 0.076 & 0.221  \\
\hline S2  & 0.001 & 0.001    & 0  & 0.008 & 0.001  & 0.010   \\
\hline S3   &  0.002 &  0.004   &  0  &  0.003  &  0  &  0.014 \\
\hline S4 & 0 & 0  &  0.001 & 0.004  &  0.001  &  0.008  \\
\hline S5  & 0.009 &  0.026 &  0.015  &  0.021   &  0.011  &   0.050 \\
\hline
\end{tabular}
\caption{The proportion of equal estimated probability for the Joint2d method with and without the algorithm selection. The column `best' represents results with the algorithm selection selection.}
\label{tab:pcs_eq_algorithm_joint2d}
\end{table}

\end{appendices}

\clearpage

\printbibliography

\end{document}